\newif\ifabstract
\abstracttrue   
\abstractfalse   

\ifabstract  

\documentstyle[twocolumn]{article}
\makeatletter
\def\section{\@startsection {section}{1}{\z@}{-3.5ex plus -1ex minus
      -.2ex}{2.3ex plus .2ex}{\large\bf}}
\makeatother

\else  

\documentstyle[11pt]{article}

\fi 
\newlength{\figboxwidth}             
\def\8{\infty}
\def\oh{\frac{1}{2}} 
 
\def\oq{\frac{1}{4}}

\def\d{\partial} 
\def\i{\imath\,} 
\def\ih{\frac{\imath}{2}\,}
\def\undertext#1{\vtop{\hbox{#1}\kern 1pt \hrule}}
\def\ra{\rightarrow}

\def\Ra{\Rightarrow}

\def\lrb#1{\left(#1\right)}
\def\O#1{O\left(#1\right)}
\def\VEV#1{\left\langle\,#1\,\right\rangle}
\def\tr{\hbox{tr}\,}

\def\pp#1{\frac{\partial}{\partial#1}}
\def\pbyp#1#2{\frac{\partial#1}{\partial#2}}
\def\ff#1{\frac{\delta}{\delta#1}}
\def\fbyf#1#2{\frac{\delta#1}{\delta#2}}

\def\br{\\ \nonumber & &}
\def\brr{\right. \\ \nonumber & &\left.}
\def\inv#1{\frac{1}{#1}}
\def\be{\begin{equation}}
\def\ee{\end{equation}}
\def\bea{\begin{eqnarray} & &}
\def\eea{\end{eqnarray}}
\def\ct#1{\cite{#1}}
\def\rf#1{(\ref{#1})}
\def\EXP#1{\exp\left(#1\right)}
\def\TEXP#1{\hat{T}\exp\left(#1\right)}

\def\MAT{{\it Mathematica }}

\def\RHS{right-hand side }
\def\COM#1#2{\left\lbrack #1 \, ,\, #2\right\rbrack}
\def\AC#1#2{\left\lbrace #1 \, , \, #2\right\rbrace}
\def\ad#1{a^{\dagger}_{#1}}
\def\SW#1{ {\cal W}\left[#1\right]}
\def\SM#1{ {\cal M}\left[#1\right]}
\def\SC#1{ {\cal C}\left[#1\right]}
\def\X#1{ {\hat{X}}_{#1}}
\def\XX#1#2{\COM{\X{#1}}{\X{#2}}}
\def\Y#1{ {\hat{Y}}_{#1}}

\def\vac#1{\VEV{ 0\left|#1\right|0}}

\def\r0{\left| 0 \right\rangle}
\def\l0{\left\langle 0 \right|}
\def\MINT#1{\int \frac{d^d #1}{(2\pi)^{d}}}
\def\P{\left| 0 \right\rangle \left\langle 0\right|}
\def\E#1{\varepsilon_{#1_1\dots #1_d}}
\def\lst#1{\left\lbrace#1\right\rbrace}
\begin{document}
\topmargin 0pt
\oddsidemargin 5mm
\headheight 0pt
\topskip 0mm

\addtolength{\baselineskip}{0.20\baselineskip}

\hfill PUPT-1655

\hfill October 1996
\begin{center}

\vspace{36pt}
{\large \bf
Hidden Symmetries of Large N QCD}

\vspace{24pt}

{\sl Alexander Migdal }

\vspace{12pt}
Department of Physics\\
and\\
Program in Applied and Computational  Mathematics\\
Princeton University\\
Princeton, NJ 08544, USA

\end{center}

\vfill

\begin{center}
{\bf Abstract}
\end{center}

\vspace{12pt}

\noindent
The new symmetry of the loop dynamics of QCD is found. This is a local
SUSY $\delta s = \theta \beta(s), \delta \theta = \beta(s)$ of the
superloop field $X_\mu(s,\theta) = x_\mu(s) + \theta \psi_\mu(s)$. The
remarkable thing is, there is no einbein-gravitino on this theory,
which makes it a 1D topological supergravity, or locally SUSY quantum
mechanics. Using this symmetry, we derive the large $N_c$ loop
equation in momentum superloop space.  Introducing as before the
position operator $\X{\mu}$ we argue that the superloop equation is
equivalent to invariance of correlation functions of products of these
operators with respect to certain quadrilinear transformation.  As a
consequence of this nonlinear symmetry, the coefficients of the
Voiculesku expansion of the position operator satisfy recurrent
equations. The generators of this symmetry are involved in the
glueball spectrum, as it follows from the loop-loop correlation
function equation. The correlation functions of external flavor currents in the
background of constant flavor gauge field with finite density of
topological charge in the chiral limit are also considered. We
represent them as certain finite dimensional integrals in superspace,
involving the Greens function, which is expressed in terms of
$\X{\mu}$. The expansion in powers of external momenta $k_1\dots k_n$
is {\em calculable} in terms of the above universal numbers in the
Voiculesku expansion.

\vspace{24pt}

\vfill
\newpage

\tableofcontents

\section{Introduction}

Recently \ct{Mig94}, some progress was achieved in a long-standing
problem of string representation of large $N$ QCD.  We introduced the
position operator $\X{\mu}$ of the endpoint of the QCD string and
applied the momentum loop equation \ct{Mig86,BVM} (MLE) to get some
relations between the planar connected moments of vacuum expectation
values of this operator.

The non-commutative probability theory \ct{Voiculesku, Cvitanovic,
Gross94} was then used to build the Fock space and operator expansion
of $\X{\mu}$ in powers of the creation operator $\ad{\mu}$ which obeys
the Cuntz algebra $ a_\mu \ad{\nu} = \delta_{\mu\nu} $.  The Fock
space consists of words $ \ad{\mu_1}\ad{\mu_2} \dots \ad{\mu_n} \r0 $,
where the vacuum $\r0 $ is annihilated by $a_\mu$.

There are several problems with this approach. One problem is that
some coefficients of this operator expansion
\be
\label{OPE}
\X{\mu} = a_\mu + A \ad{\mu} + B \ad{\nu} \ad{\mu} \ad{\nu} + C
\lrb{\ad{\nu} \ad{\nu} \ad{\mu} + \ad{\mu} \ad{\nu} \ad{\nu}} + \dots.
\ee
are left undetermined. Apparently, some information is still
missing.  Another problem is that there seems to be no direct
relation between these coefficients and the QCD observables.

As for the first problem, it was known since the first papers on the
loop equation \ct{MM,Mig83} that one should restrict oneself to the
class of so-called Stokes type functionals to reconstruct the Wilson loop 
from the loop equation. In those papers it was shown that iterations of the loop 
equation within this class of functionals reproduce the Faddeev-Popov 
perturbation theory including ghost loops.

The Stokes type functionals by definition satisfy the Mandelstam relation
\be
\label{Mand}
\ff{x_\mu(s)} = x'_\nu(s) \pp{\sigma_{\mu\nu}(s)},
\ee
where the area derivative $\pp{\sigma_{\mu\nu}(s)}$ is a skew 
symmetric tensor operator satisfying the Bianchi identity
\be
\d_\alpha(s) \pp{\sigma_{\beta\gamma}(s)} + {\rm cyclic} = 0.
\ee
The derivative $\d_\mu(s)$ is defined as
\be
\d_\mu(s) = \int_{s-0}^{s+0} \, d t\,\ff{x_\mu(t)}.
\ee

These relations were not taken into account in our previous paper
\ct{Mig94}. Also, the loop equation used in that paper was not the
most general one. It corresponded to the original vector loop equation
\ct{MM} multiplied by the tangent vector $x'_\nu(s) $ and integrated
around the loop. Normal component terms were eliminated by this
projection. For example, in two dimensions, the loop averages were
computed \ct{KazKos} using the normal component of the loop
equation. The tangent components in this case were identically
satisfied for the area dependent Ansatz.

The second problem  is more serious. The operator $\X{}$ as defined is
apparently not an observable, so it may be infinite or zero in the
local limit of the theory. 
 
In this paper we fix both of these problems at the same time.  We
choose the superloop formalism (see, e.g. \ct{Pol87,Tseytlin}) which
allows one to impose Mandelstam and Bianchi relations as local
supersymmetry relations.  We derive the large $N$ superloop equation
in momentum space. 
 
The SuperFourier transformation we use here, is free of the zero point
fluctuation problems, which were a serious obstacle to the old
momentum loop dynamics. These fluctuations cancel by themselves,
without any Wiener measure, by virtue of supersymmetry.

Thus, there are no more  hidden infinite renormalization factors in
our momentum superloops. The corresponding supersymmetric
generalization of the position operator of the string obeys a different
set of recurrent equations, which are  more restrictive than before.

These relations are based on the important observation that there is a
local  SUSY in the loop dynamics. The local SUSY is valid in the large
$N_c$ loop equations as well as in the relations for these observables
in the chiral limit. We are able to explicitly find all the
superspace-dependent quantities (they reduce to the SUSY $\Theta $
functions).

The physical interpretation of these relations is found. They represent
the {\em symmetry} relations for an operator $\X{}$ correlation
functions. One can perform certain nonlinear transformation of $\X{}$
operator with two vector parameters. This symmetry is {\em completely}
equivalent to the momentum loop equation.  

The non-commuting vector generators of this symmetry act as ``QCD
string'' endpoint translations. Their eigenfunctions must have fixed
spatial momentum, which condition yields the glueball spectrum
equation. We derive this condition from the glueball superloop
equation.

The superloop in momentum space has direct physical meaning in the
chiral limit of QCD. It is related to the correlation functions of
external flavor currents.  The local supermomentum is expressed in
terms of the set of external momenta entering the quark loop.

One set of correlation functions involves the chiral anomaly (the
periodic boundary conditions for the Fermi component of the
superpath). Those are expected to be completely dominated by the
perturbative contribution. The argument is based on the index
theorem. However, in presence of non-smooth fluctuations of the gluon
field this perturbative argument may fail. We relate the coefficients
in the low energy expansion of these correlation functions to the
coefficients of the Voiculesku expansion. 

Another set of calculable observables is given by the ordinary
correlation functions of external flavor currents taken in the
background of external field with finite density of topological charge
(the anti-periodic boundary conditions for the Fermi component of the
superpath). The coefficients of the low energy expansion are also
related to the coefficients of the Voiculesku expansion. We derive an
explicit set of operator equations in presence of external field.

The resulting theory can be reformulated as the theory of the quark
superfield propagating around the  superloop, rather than target
space. The corresponding propagator can be found exactly in terms of
the quark position operator $\X{\mu}$. This resembles the string
theory, but we do not see any precise correspondence. 

In order to handle the chiral symmetry breaking in absence of the
external density of topological charge we included the quark mass
term. This resulted in extra term in the quark effective action on a
superloop. This term appears to be a quartic interaction, rather that
a mass term. This superfield theory is essentially one dimensional, so
it will hopefully present no new problems. 

\section{Vacuum Energy and Superloops}

Let us present our superloop formulation of QCD. It starts with the
formula (derived in Appendix A) for the quark loop energy in the
presence of color and flavor fields
\bea {\cal E} = const -\oh \tr
\log \lrb{m_0^2 + \lrb{\i\gamma_\mu\nabla_\mu}^2} = \br \oh\int_0^{\8}
d T T^{\alpha-1}\lrb{1 + \alpha \ln T}e^{-T m_0^2} \tr
\EXP{-T\lrb{\i\gamma_\mu\nabla_\mu}^2}.
\eea
In this and subsequent formulas the limit $\alpha \ra +0$ is
understood. This $\alpha$ prescription is equivalent to the $\zeta$
regularization and it goes along with dimensional regularization.

The covariant derivative operator involves all the gauge
potentials $\nabla_\mu = \d_\mu + A_\mu + B_\mu $, where  $A_\mu(x)$
is the quantum gluon field, and $B_\mu(x)$ is the external flavor field.

The  SUSY follows from the  effective Hamiltonian
\bea
 \lrb{\i\gamma_\mu\nabla_\mu}^2 = \AC{ Q_L}{Q_R},\br
 Q_{L,R} = \lrb{\i\gamma_\mu\nabla_\mu}\frac{1\pm \gamma_5}{2}.
\eea
Therefore a Dirac operator generates the (global) SUSY
transformations. This is the well-known SUSY Quantum Mechanics. 

The path integral representation can be formulated in terms of  the
super trajectory $x_\mu(s), \psi_\mu(s) $.  The Fermi coordinates
$\psi_\mu(s)$ represent the the matrices $\gamma_\mu \sqrt{T}$. 

It is most convenient to use the superspace $ S = (s,\theta)$ with the 
superfield
\be
X_\mu(S) = x_\mu(s) + \theta \psi_\mu(s).
\ee
with boundary conditions of (anti)periodicity
\bea
x_\mu(1) = x_\mu(0),
\psi_\mu(1) = -\psi_\mu(0).
\eea
The negative sign for $\psi_\mu$ is needed to make the traces of
$\gamma$ matrices cyclic symmetric in spite of anti-commutation of
$\psi_\mu$. Moving the left $\psi_\mu$ to the right all the way
through remaining (odd) number of $\psi$ would produce the minus sign,
unless we compensate it by the extra sign flip at $s=1$. 

The action is given by an integral over superloop 
\be
{\cal I} = \int_0^1 ds \int d\theta \, {\cal L}(S)
\ee
with  the  super Lagrangian (in our normalization of proper time and 
$\psi$ fields)
\bea
{\cal L}(S) = {\cal L}^{kin}(S)  \oplus D X_\mu(S) A_\mu(X(S))  
\oplus D X_\mu(S)B_\mu(X(S)),\br
{\cal L}^{kin}(S) = -\inv{4T} \, D X_\mu(S) D^2 X_\mu(S)
\eea
Here
\be
D = \pp{\theta} + \theta \pp{s},
\ee
is a covariant derivative in superspace. Its square is the time derivative
\be
D^2 = \pp{s}.
\ee

The direct sum of the color and flavor matrices in understood in the
sense of taking the ordered exponential of the action in the
(super)path integral 
\bea
\int {\cal D} X \TEXP{ {\cal I}} = \int {\cal D} X \EXP{\int_0^1 ds
\int d\theta \, {\cal L}^{kin}(S)} \br 
\TEXP{ \int_0^1 ds \int d\theta \,D X_\mu(S) \lrb{A_\mu(X(S))  \oplus 
B_\mu(X(S)}}
\eea
which is the  non-Abelian generalization of the Feynman's path
integral. One may prove that the results are equivalent to the more
traditional approach with the ordinary exponential of a second
quantized action.  

Conventional formalism shifts the momentum by $\i p_\mu \Ra \i p_\mu \oplus
A_\mu \oplus B_\mu$ after which the gauge fields go to kinetic terms
in the Lagrangian, which is inconvenient. So we undid this shift,
coming back to original ideas of Feynman, who started with the
(Abelian) Wilson loop as an amplitude for electron propagation. In our
previous paper \ct{Mig86} we discussed these issues in detail.  

\section{Superloop Kinematics}

In our case it is implied that the gauge fields have the superfield as
their argument, which means that first derivative terms are also
included. The (global) SUSY generalization of the the ordered product
is also known \ct{Tseytlin}. 
One introduces a SUSY generalization of the distance
\be
L(S,S') = s' - s + \theta\theta',
\ee
and the corresponding theta function
\be
\label{THETA}
\Theta(S,S') = \Theta(L(S,S')) = \Theta(s'-s)  +\theta\theta'\delta(s'-s).
\ee
The covariant derivative $ D = \pp{\theta} +\theta \pp{s}$ of this
$\Theta$ function reduces to the SUSY $\delta$ function 
\be
D \Theta(S,S') = (\theta'-\theta) \delta(s'-s) \equiv \delta(S'-S).
\ee
This $\delta$ function is a Fermi rather than Bose element, and it is an 
{\em odd} function
\bea
\delta(S) = -\delta(-S),\br
\delta(0) = 0.
\eea
The last property is very convenient for computations of the SUSY graphs.

The generalized $\Theta$ function satisfies the usual property
\bea
\Theta(A,B) + \Theta(B,A) = 1,
\eea
from which it follows that $\Theta(A,A) = \oh$.

We shall also need the generalization of the finite range integral
\be
\int_A^B d S \equiv \int_{-\8}^{\8} d s \int d \theta \Theta(A,S) 
\Theta(S,B),
\ee
as well as the multiple ordered integral
\be
\int_A^B d^n S \equiv \int_{-\8}^{\8} d s_1 \int d \theta_1 \dots
\int_{-\8}^{\8} d s_n \int d \theta_n \Theta(A,S_1)
\Theta(S_1,S_2)\dots \Theta(S_{n-1},S_n) \Theta(S_n,B). 
\ee 

The following superfield differential forms will play an important role
\bea
\label{FORMS}
dX_\mu(S) = d s d \theta D X_\mu(S),\br
d^n X_{\lst{\mu}}(\lst{S}) = d X_{\mu_1}(S_1) \dots d X_{\mu_n}(S_n).
\eea
By construction these forms are globally SUSY, later we prove that they 
are locally SUSY as well.
We are using the curly brackets $\lst{ S}$ for lists of
indexes or arguments and the tensor power notation 
\be
F^n_{\lst{ \mu}}(\lst{ S}) \equiv \prod_{i=1}^n F_{\mu_i}(S_i).
\ee 

The superfield differential $1$-form satisfies the usual integration 
identities
\bea
\int_A^B d F(S) = F(B)-F(A),\br
d (F(S)  G(S)) = d F(S) G(S) + F(S) d G(S),
\eea
which allow integration by parts. The finite range integrals are additive
\be
\int_A^B d F(S) G(S) = \int_A^C d F(S) G(S) + \int_C^B d F(S) G(S).
\ee

The integrals of multiple forms satisfy the composition law
\bea
\label{COMP}
\int_A^B d  P_\lambda(S) \int_A^S d^m P_{\lst{ \mu}}(\lst{U}) \int_S^B d^n 
P_{\lst{\nu}}(\lst{ V})=  
\int_A^B d^{m + n +1} P_{\lst{\rho}}(\lst{ W}),
\eea
with the joined list of indexes
\be
\lst{ \rho} = \lst{ \lst{ \mu},\lambda,\lst{ \nu}}.
\ee
as well as the multiplication laws, the simplest of which is
\bea
\label{MULT}
\int_A^B d  P_\lambda(U) \int_A^B d^n  P_{\lst{ \mu}}(\lst{ V}) =
\sum_{l=0}^{n} \int_A^B d^{n+1}  P_{\lst{ \nu(l)}}(\lst{ S}),
\eea
where 
\be
\lst{ \nu(l)} = \lst{ \mu_1,\dots\mu_{l},\lambda,\mu_{l+1}\dots\mu_n}.
\ee
In other words the index $\lambda$ is inserted at the position $l+1$ in 
the list of $\mu$ indexes.

Both of these laws follow from the local SUSY which we prove in the
next Section and the supersymmetrization theorem which we prove in
Appendix C. 
 
Using these forms, we can define the SUSY path exponential with
initial point $A = (a,\theta_a)$ and final point $B = (b,\theta_b)$ 
\be
{\cal U}_A^B\lbrack  A_\mu,X_\mu\rbrack \equiv \TEXP{\int_A^B d X_\mu(S) 
A_\mu(X(S)) } \equiv 
\sum_{n=0}^\8 \int_A^B d^n X_{\lst{ \mu}}(\lst{ S})
A^n_{\lst{ \mu}}\lrb{\lst{ X(S)}}. 
\ee

One may verify the multiplicativity of the path exponent
\be
{\cal U}_A^B\lbrack  A_\mu,X_\mu\rbrack = {\cal U}_A^S\lbrack
A_\mu,X_\mu\rbrack{\cal U}_S^B\lbrack  A_\mu,X_\mu\rbrack, 
\ee
its differential with respect to the endpoints
\bea
d {\cal U}_A^S\lbrack  A_\mu,X_\mu\rbrack  = \delta(S-A) + {\cal 
U}_A^S\lbrack  A_\mu,X_\mu\rbrack d X_\mu(S) A_\mu(X(S)),\br
d {\cal U}_S^B\lbrack  A_\mu,X_\mu\rbrack  = -\delta(S-B) - d X_\mu(S) 
A_\mu(X(S)){\cal U}_S^B\lbrack  A_\mu,X_\mu\rbrack,
\eea
as well as the usual formula for its variation with respect to the gauge 
potential
\be
\delta_{A_\mu} {\cal U}_A^B\lbrack  A_\mu,X_\mu\rbrack = \int_A^B d
X_\mu(S){\cal U}_A^S\lbrack  A_\mu,X_\mu\rbrack\delta A_\mu(X(S))
{\cal U}_S^B\lbrack  A_\mu,X_\mu\rbrack. 
\ee
In particular, for the gauge variation $ \delta_{gauge} A_\mu(X) = 
\nabla_\mu \alpha(X)$ (note that the gauge parameter $\alpha(X)$ having 
the superfield $X(S)$ as its argument)
\be
\delta_{gauge} {\cal U}_A^B\lbrack  A_\mu,X_\mu\rbrack = \int_A^B d
X_\mu(S){\cal U}_A^S\lbrack  A_\mu,X_\mu\rbrack\nabla_\mu \alpha(X(S))
{\cal U}_S^B\lbrack  A_\mu,X_\mu\rbrack. 
\ee
Using above endpoint derivatives we reduce the integral to the total 
derivative, up to the $\delta$ terms, which yields
\be
\delta_{gauge} {\cal U}_A^B\lbrack  A_\mu,X_\mu\rbrack = {\cal 
U}_A^B\lbrack  A_\mu,X_\mu\rbrack \alpha(X(B)) - \alpha(X(A)) {\cal 
U}_A^B\lbrack  A_\mu,X_\mu\rbrack .
\ee

We got something new here. The trace of this variation will vanish 
provided the superloop is closed, i.e.  $ X(B) = X(A)$. The Bose and Fermi 
parts by themselves could be opened. The gap in Bose part $x(b)-x(a) = 
-\theta_b \psi(b) + \theta_a \psi(a)$ represent the nilpotent even element of the 
Grassmann algebra. Its higher tensor products $ 
(x(b)-x(a))_{\mu_1}\dots (x(b)-x(a))_{\mu_n}$ vanish at $n>2$. This is 
less restrictive than the ordinary gauge invariance which would demand $ 
n>0$. 

We are in a position to define  the superloop 
\bea
\label{TEXP}
\SW{X}_A^B = \inv{N_c}\VEV{ \tr {\cal U}_A^B\lbrack  
A_\mu,X_\mu\rbrack}_{A_\mu}.
\eea
In usual notation the SWLoop reads
\be
\SW{X}_0^1 = 
\inv{N_c} \tr \TEXP{\int_0^1 d s \lrb{ A_\mu(x) x'_\mu + \oh
    \psi_\mu\psi_\nu F_{\mu\nu}(x)}} \equiv \SW{x,\psi}.
\ee
This formula was first obtained in \ct{Luis,Dan} for the same purpose
of supersymmetric description of spin $\oh$ particle.  In case of the 
Grassmann variables at the end points $A,B$, there will be extra terms. 
Say, in Abelian case there will be extra factors $ \EXP{\pm \theta 
\psi_\mu A_\mu(x)} $ at the endpoints.

The relation between this formula and the one with superfield is quite 
amazing. 
The $\psi\psi \d A$  part of the $\psi\psi F$ terms comes from the
expansion of the  potential in the action, and the commutator terms $
\psi\psi A A $ come  from the $\theta_i\theta_j \delta(s_i-s_j)$ term
in the SUSY definition of the the ordered product \ct{Tseytlin}. The
SUSY and the non-Abelian gauge invariance work in close collaboration,
which is quite common, as we shall see. 

The vacuum energy can be written in terms of superfields as follows 
\bea
\label{VACEN}
{\cal E} = 
\frac{ N_c}{2 }\int_0^\8 d T T^{\alpha-1}\lrb{1 + \alpha \ln T}e^{-T 
m_0^2}\br
\int {\cal D}X \EXP{ -\frac{1}{4T}\int_0^1 d X_\mu(S) D^2 X_\mu(S)}
\tr \TEXP{ \int_0^1 d X_\mu(S) B_\mu(X(S)) }\, \SW{X}_0^1.
\eea
The ordinary perturbation theory corresponds to expansion in both
flavor and color gauge fields and taking the Gaussian path integrals
term by term. The role of the Grassmann part $\theta \psi_\mu(s)$ of
the path $X_\mu(S)$ is to get correct numerators of the Feynman
diagrams, coming from the Dirac traces in usual language.  

This follows from the explicit form of the kinetic energy
\be
\frac{1}{4T}\int_0^1 d  X_\mu(S) D^2 X_\mu(S)  = \frac{1}{4T}\int_0^1
ds \lrb{x'_\mu(s)x'_\mu(s) + \psi'_\mu(s)\psi_\mu(s)}. 
\ee
One could rescale $s \Ra T \, s, \psi_\mu \Ra  \sqrt{T}
\,\psi_\mu $, which only affects this term, as the rest of the terms
are parametric invariant. Then the kinetic energy will take a familiar
form, which corresponds to the usual free particle. The $\psi$ part
after quantization  reproduces the usual Clifford algebra of Dirac
matrices. The extra factors of mass in the local measure, coming from
rescaling of $\psi_\mu$, are just the correct factors needed to
normalize the Wiener measure for $x_\mu(s)$. 
 
We mentioned this only to make our path integral look more
familiar. For our purposes this transformation is not needed, in fact,
it would only make the formulas look more complex than they actually
are.  

For example, the Euler-Heisenberg effective Lagrangian \ct{Zuber} in
homegeneous external field $B_\nu(x) = \ih B_{\mu\nu} x_\mu$ is
computed in just few lines in Appendix B. 

\section{Correlation Functions of Vector Currents}

Let us expand the vacuum energy in powers of the external field.  The
expansion coefficients are related to the correlation functions of the
vector flavour currents. 

We use the Fourier integrals
\be
B_\mu(X) = \MINT{k} B_\mu(k) \EXP{\i k_\mu X_\mu},
\ee
which also define the SUSY extension to $X = X(S)$. In the $n$-th
order we shall have the product of $n$ exponentials, which can be
written in terms of the supermomentum 
\be
\EXP{\i \sum_i k_\mu^{i} X_\mu(S_i)} = \EXP{\i \int_0^1 d K_\mu(S) 
X_\mu(S) }=
\EXP{-\i \int_0^1 d X_\mu(S) K_\mu(S) },
\ee
with the supermomentum
\bea
K_\mu(S) = \sum_i k_\mu^i \Theta(S_i,S),\br
d K_\mu(S) = \sum_i k_\mu^i d S \delta(S-S_i).
\eea

We need  the standard SUSY functional derivative for a superfield
$P_\mu(s) = p_\mu(s) + \theta \varphi_\mu(s) $  
\be
\ff{P_\mu(S)} = \ff{\varphi_\mu(s)} +\theta  \ff{p_\mu(s)},
\ee
which satisfies the identity
\be
\ff{P_\mu(S)} P_\nu(S') =  \delta_{\mu\nu} \delta(S-S').
\ee
In the same way, as we introduced the superfield $1$-form $ d
X_\mu(S)$ we could introduce an invariant superfield gradient $1$-form

\be
\pp{P_\mu(S)} \equiv d S \ff{P_\mu(S)}.
\ee
The motivation for this notation is the following identity
\be
\pp{P_\mu(S)} \int_A^B d X_\nu(U) P_\nu(U)  =  d X_\nu(S),
\ee
which means that this gradient form acts on the integral of
differential form just as it does in Bosonic case. 

The vacuum energy takes the form
\be
\label{EP}
{\cal E} = 
\sum_{n=0}^\8 \i^n \MINT{k_1}\dots \MINT{k_n} \,\delta^d\lrb{\sum_i
k_i}\,B^{a_1}_{\mu_1}(k_1) \dots
B^{a_n}_{\mu_n}(k_n)\,\VEV{J^{a_1}_{\mu_1}\dots
J^{a_n}_{\mu_n}}\lrb{k_1\dots k_n}, 
\ee
with the following planar connected correlation functions of flavour 
currents
\bea
\label{GENERAL}
\VEV{J^{a_1}_{\mu_1}\dots J^{a_n}_{\mu_n}}\lrb{k_1\dots k_n} = \br
\frac{ N_c}{2 }\,
\tr\lrb{\tau_{a_1}\dots\tau_{a_n}}\,Z\lrb{\frac{1}{4}\int_0^1
\pp{K_\mu(S)} D \ff{K_\mu(S)}} \,  
\int_0^1   \pbyp{\SM{K}_0^1}{K_{\mu_1}(S_1)\dots \d K_{\mu_n}(S_n)},
\eea
where we introduced the following operator function
\be
Z(\hat{L}) = \lim_{\alpha \ra 0}\int_0^\8 d T T^{\alpha-1}\lrb{1 +
\alpha \ln T}\EXP{-T m_0^2 - \frac{\hat{L}}{T}}. 
\ee
This is a certain Bessel function of its operator argument.

The loop functional 
\be
\label{FOURIER}
\SM{P}_A^B = \int {\cal D} X \delta^d(X(A))\EXP{\i \int_A^B d P_\mu(S) 
X_\mu(S) } \SW{X}_A^B;
\ee
is a Fourier transform in superspace.

\section{Chiral Anomaly}

There is a simpler observable, which is not influenced by quark
kinetic energy. This is the chiral anomaly 
\be 
I[B] = \VEV{\tr \gamma_5 \EXP{- T \lrb{\i\gamma_\mu\nabla_\mu}^2}}.
\ee
as a functional of the external flavour field $B$. The famous index
theorem states that it is $T$-independent: due to the $\gamma_5$
symmetry of the Dirac operator, all the finite eigenvalues drop from
the trace, so that only the $T$- independent contribution of the zero
eigenvalies remain. In  absense of quantum fields, this allows to
compute it, taking the limit $T\ra 0$ and using the WKB
expansion. This leads to the the Grassmann integral over the zero mode
\be
I_{cl}[B] \propto V \int d^d \psi^0 \tr\TEXP{ \oh B_{\mu\nu} 
\psi^0_\mu\psi^0_\nu} \propto 
V \E{\mu} \tr B_{\mu_1\mu_2}\dots B_{\mu_{d-1}\mu_d}.
\ee
In case of abelian field this is the pfaffian $ \sqrt{\det \i B}$.
 
The quantum chiral anomaly, $I[B]$ according to the same index theorem,
is simply the vacuum average of the classical chiral anomaly of the direct sum of the two gauge 
fields $\VEV{I_{cl}[B\oplus A]}$ computed in absence of
the $B$ field. 
\be
I[B] = \VEV{I_{cl}[B\oplus A]} = N_c I_{cl}[B] + N_f \VEV{I_{cl}[A]}.
\ee
The last term vanishes by parity, so that  the chiral anomaly
stays the same as in the free quark theory. 

Let us now relate this vacuum average to our momentum
loops. We have to repeat the steps of the previous section, with two
changes. First, we choose the different boundary conditions. Now both
$x(s)$ and $\psi(s)$ are periodic, to compensate for the
$\gamma_5$. After that we can drop the kinetic energy term, according
to the index theorem. 

We find the following formula
\be
\label{ANOM}
 N_c I_{cl}[B] = \frac{ N_c}{2 } \int_{periodic} {\cal D} X \TEXP{\int_0^1 d 
X_\mu(S)B_\mu(X(S))} \SW{X},
\ee
which provides an important normalization condition for the SMloop 
considered below.

In particular, consider the case  of constant non-abelian field
$B\mu(X) = const$. This corresponds to constant field strength $
B_{\mu\nu} = \COM{B_\mu}{B_\nu}$. The chiral anomaly reads
\be
I_{cl}[B] \propto V \E{\mu} \tr B_{\mu_1\mu_2}\dots B_{\mu_{d-1}\mu_d}
\propto V \E{\mu} \tr B_{\mu_1}B_{\mu_2} \dots B_{\mu_d}.
\ee
Comparing this  with above path integral we get the sum rule
\be
\int_{periodic} {\cal D} X  \int_0^1
d^d\X{\lbrace\mu\rbrace}\lrb{\lbrace S \rbrace}\SW{X} \propto \E{\mu}.
\ee
This fixes the coefficients in front of $ K_{\mu_1}\dots K_{\mu_d}$ in
Fourier transform of $\SW{X}$ with periodic boundary conditions.

\section{Local SUSY}

Let us study the symmetries of the SWloop. It is invariant under 
parametric transformations
\be
\delta x_\mu(s) = \alpha(s) x'_\mu(s),\; 
\delta\psi_\mu(s) = \alpha(s) \psi'_\mu(s) + \oh \alpha'(s) \psi_\mu(s),
\ee
which is straightforward to verify. In addition it is invariant  under
local SUSY transformations 
\be
\delta x_\mu(s) = \beta(s) \psi_\mu(s),\;
\delta \psi_\mu(s) = \beta(s) x'_\mu(s).
\ee
with some Grassmann variable $\beta(s)$. Together these transformations
define the reparametrization of the superfield,
\be
\delta X_\mu = X_\mu(s + \delta s, \theta + \delta \theta) -
X_\mu(s,\theta) = \lrb{\delta s \pp{s} + \delta \theta \pp{\theta}} 
X_\mu(s,\theta),
\ee
with
\be
\label{DST}
\delta s = \alpha(s) + \theta \beta(s),\;
\delta \theta =\oh \alpha'(s)\theta  + \beta(s).
\ee

From now on, we shall always mean the local rather than the global SUSY.
In Abelian theory this is just the formal symmetry of the gauge field
part of the action. One may verify that the variation of the $A-$
term in a Hamiltonian after integration by parts cancels the variation
of the $F-$ term. In the last variation  the extra terms with
gradients of $F_{\mu\nu}$ add up to zero by virtue of the Bianchi
identity 
\be
\psi_\mu\psi_\nu\psi_\lambda \d_\lambda F_{\mu\nu} = 0
\ee

In the non-Abelian theory the same can be proven for the the ordered
exponent. In the superfield language the SUSY of the action follows
from the opposite transformation laws of the integration measure and
covariant derivative 
\be
\delta d S =  \lrb{\frac{\d(\delta s, \delta \theta)}{\d(s,\theta)} -
1} d S = \lrb{\pbyp{\delta s}{s} + \pbyp{\delta \theta}{\theta}} d S =
\lrb{\oh \alpha'(s) + \theta \beta'(s)} d S, 
\ee
and
\be
\delta D = \delta \pp{\theta} + \delta \theta \pp{s} + \theta \delta 
\pp{s} =
-\lrb{\oh \alpha'(s) + \theta \beta'(s)} D.
\ee
so that the differential $ d  = d S D $  stays invariant.

It remains to check  the $\Theta $ function \rf{THETA}. The parametric
invariance is obvious, as for  the local SUSY, it can be verified as
follows 
\bea
\delta\Theta(S,S') = \delta L(S,S') \delta(s-s' + \theta \theta') =\br
(\theta +\theta')(\beta(s')-\beta(s))\delta(s-s' + \theta \theta') =\br
(\theta +\theta')(\beta(s')-\beta(s))\delta(s-s') = 0.
\eea
It is interesting that Andreev and Tseytlin \ct{Tseytlin}, who
introduced this SUSY $\Theta$ function, did not notice its local SUSY,
as they were working in a superstring theory, where this local
symmetry was broken down to a global one. 

From the point of view of the 1D supergravity this local SUSY
corresponds to the fact that the Wilson loop does not depend on the
metric in 1D superspace, described by the einbein-gravitino
superfield. Thus it measures topology in this space, which is rather
simple. 

It is quite instructive to verify the local SUSY directly, using the
components of the superfield. We refer the reader to the review paper
\ct{Mig83} where the loop  kinematics and dynamics was discussed in
great detail. Using the methods of that paper one can prove the SUSY
of the non-Abelian SWLoop as follows. 

From the point of view of loop dynamics the SWLoop is not an 
independent functional. It relates to the Wilson loop by the linear 
operator
\be
\label{SWW}
\SW{x,\psi}_0^1 = \TEXP{\int_0^1 d s \oh \psi_\mu \psi_\nu
\pp{\sigma_{\mu\nu}}} W\lbrack  x\rbrack_0^1. 
\ee

Using this representation, the functional derivatives  can be readily 
computed
\bea
\fbyf{\SW{x,\psi}_0^1}{\psi_\mu(s)} = \psi_\nu(s) 
\fbyf{\SW{x,\psi}_0^1}{\sigma_{\mu\nu}}(s)
,\br
\fbyf{\SW{x,\psi}_0^1}{x_\mu(s)} = \lrb{x'_\beta(s)\delta_{\alpha\mu}
+ \oh \psi_\alpha(s) \psi_\beta(s) \d_\mu(s) 
}\fbyf{\SW{x,\psi}_0^1}{\sigma_{\alpha\beta}}(s).
\eea
It is implied that the left functional derivative is used for the 
Grassmann variable $\psi$.

Multiplying the last relation by $\psi_\mu(s)$ and using the non-Abelian
Bianchi identity \ct{Mig83}, we get 
\be
\label{SUSY}
\psi_\mu(s) \fbyf{\SW{x,\psi}_0^1}{x_\mu(s)} =  x'_\nu(s) \psi_\mu(s) 
\fbyf{\SW{x,\psi}_0^1}{\sigma_{\mu\nu}}(s) = -x'_\nu(s) 
\fbyf{\SW{x,\psi}_0^1}{\psi_\nu}(s),
\ee
which is a differential form of the SUSY relation.

This local supersymmetry is a genuine symmetry of the loop
dynamics in the pseudoscalar sector. It does not require any external
einbein-gravitino fields, which were needed to supersymmetrize the
massive spin $\oh$ particle. In other words, we are dealing with
topological SUSY Quantum Mechanics. 

\section{SuperLoop Equation}

The SWLoop equation is obtained as a SUSY extension\footnote{I am
indebted to Sasha Polyakov for the advice to switch to superfields
here.} of the usual loop equation \ct{MM,Mig83}. It looks exactly the
same as the ordinary loop equation, with a superfield $X_\mu(S)$ in
place of the ordinary coordinate $x_\mu(s)$, 
\be
0 = \int_0^1 d S \lrb{\fbyf{ \SW{X}_0^1}{X_\mu(0)\delta X_\mu(S)} +
N_c g_0^2 \,DX_\mu(0) DX_\mu(S) \delta^d(X(S) - X(0)) \SW{X}_0^S 
\SW{X}_S^1}.
\ee

In Appendix C we derive the supersymmetrization theorem, which states
that any  locally SUSY functional of the curve $\Gamma: S =
(s(\tau),\theta(\tau))$ which vanishes identically for $\Gamma$ being
the real axis: $\theta(\tau) \equiv 0$ must vanish identically in the
whole superspace. The bottom line is, the local SUSY transformation
has one Bose and one Fermi  function of one variable, so it allows us
to move and reparametrize any curve  from the real axis anywhere in
superspace. 

This resembles the analytic continuation from the real axis to the
complex plane. As long as the Cauchy-Riemann equations are satisfied
this analytic continuation looks like a trivial replacement of the
real number by a complex number in any equation. The hidden cost is
just the Cauchy-Riemann equations which should be checked in every
case.  

One may think that the SUSY loop equation follows from the ordinary
loop equation by means of this naive SUSY extension, but this is not
true. The supersymmetrization theorem does not literally apply
here. The two arbitrary functions of the SUSY transformation are not
sufficient to eliminate the Grassmann components of $d$-vector
superfield $X_\mu(S)$. One has to rederive the equation from scratch,
using the superfields instead of coordinates.

Let us examine the linear term in components
\bea
 \int_0^1 d s \fbyf{\SW{X}_0^1}{\psi_\mu(0)\delta x_\mu(s)}  =
-\d_\mu(0) \ff{\psi_\mu(0)} \SW{X}_0^1= -
\psi_\nu(0)\nabla_\mu F_{\mu\nu}(x(0))\otimes\SW{X}_0^1.
\eea
with the  colour matrix $ \nabla_\mu F_{\mu\nu}(x(0))$ being 
inserted in the beginning of the the ordered product. 

According to the quantum YM equations, this can be replaced by
\be
-\psi_\nu(0)g_0^2 \ff{A_\nu(x(0))}\otimes\SW{X}_0^1
\ee
with the same agreement about insertion inside the the ordered product.
The variation of the path the ordered exponential goes as follows
(each step can be proven using the above Taylor expansion with SUSY
$\Theta$ functions) 
\bea
\ff{A_\nu(x(0))}\otimes \TEXP{\int_0^1 d X_\mu(S) A_\mu(X(S)) } = \br
\int_0^1 d  X_\mu(S) \delta^d(X(S)-x(0))\br
\TEXP{\int_0^S d  X_\mu(U)A_\mu(X(U)) }\otimes \TEXP{\int_S^1 d X_\mu(V) 
A_\mu(X(V)) }.
\eea
The $\delta^d$ function here is the usual delta function in $d$
dimensional Euclidean space, with argument replaced by the
superfield. It should be understood in the sense of a Fourier integral

\bea
A_\mu(X(S)) = \MINT{k} A_\mu(k)\EXP{\i k_\mu X_\mu(S)},\br
\fbyf{A_\mu(X(S))}{A_\nu(x(0))} = \int d^d q \EXP{-\i q_\mu x_\mu(0)}
\fbyf{A_\mu(X(S))}{A_\nu(q)} = \br
\MINT{k} \int d^d q \delta^d(k-q)\EXP{\i k_\mu X_\mu(S)}\EXP{-\i q_\mu 
x_\mu(0)}=\br
\MINT{k} \EXP{\i k_\mu( X_\mu(S) - x_\mu(0))} \equiv \delta^d(X(S)-x(0)).
\eea
Recalling that $\psi_\nu(0) = D X_\nu(0)$ and $ x_\nu(0) = X_\nu(0)$
we get the above superloop equation. 

Let us see how this works in components. The integral over the
$\theta$ variable in $\int_0^1 d X_\mu(S)$ picks up the first
derivative in the superfield expansion.  This is equivalent to
applying the operator $ \psi_\mu(s) \d_\mu(s) $ to the $\delta^d(X(S)
- X(0))$ function as well as to the SWLoops $ \SW{X}_0^S
\SW{X}_S^1$. In  the latter case these are the covariant derivatives
$\d_\mu(s)$.  

Like in a SWLoop itself, the various parts of the covariant derivative
come from various places, supported by local SUSY. The $\pp{x} $ terms
in covariant derivatives come from expansion of the $X$ superfield,
and the $A_\mu$ terms come from the expansions of the $\Theta$
functions in the the ordered product (the last $\Theta$ function
$\Theta(S_n,S)$ in the $\SW{X}_{0S}$  and the first one $
\Theta(S,S_1)$ in the $\SW{X}_{S1} $ are differentiated). We get
simply 
\be
x_\mu'(s) + \psi_\nu(s) \psi_\mu(s) \d_\nu(s).
\ee
as an operator applied to the rest of the factors. 

The same expression could be obtained by an honest variation of the
SWloop. In superfields everything looks so simple, but in fact all the
subtleties of the loop dynamics are automatically taken care of.  

Let us discuss the boundary conditions. The initial loop $\SW{X}_0^1$
was periodic in $x(s)$ but the boundary values of $\psi(s)$ were left
free. The loops $\SW{X}_0^S, \SW{X}_S^1$ in the nonlinear part of
equation belong to the same class. 

The right boundary value $\psi(s)$ of the of the first part of the
path coinsides with the left boundary value for the second part, so
that we have a convolution in the loop equation. 
 
The physical loops, which enter the vacuum energy and the vacuum
topological  charge, are closed in $\psi$ with antiperiodic or
periodic boundary conditions. This can be achieved by setting $\psi(1)
= \pm \psi(0)$ and integrating over $\psi(0)$.

\section{Momentum Superloops}

Let us now go to supermomentum space.  There is one important point to 
clarify before we do it.  The natural definition would imply the periodic 
momentum $P_\mu(S)$ , $P_\mu(1) = P_\mu(0)$.  However, the loop equation 
does not close in the space of periodic momentum loops.

This issue was discussed before \ct{Mig86,Mig94}. In general, one
could introduce some gaps $\Delta P(S_i) = P(S_i +0) - P(S_i-0)$, in
which case the momentum loop equation relates the $n-$ gap functional
to bilinear superposition of $l+1$ and $ n-l+1$ functionals, with $ l
= 0,\dots n$. 

These gaps correspond to slow decrease $\inv{n}$ of the Fourier
harmonics of $p_\mu(s)$, which is an unpleasant complication of
momemtum loop dynamics. As we see it now, this complication can be
avoided. Take {\em non-periodic} functions, $P_\mu(1) \ne P_\mu(0) $
which are otherwise smooth. The derivative $ D P_\mu(S) $ could be
finite everywhere including the endpoints. However, the endpoint
derivatives are different $DP_\mu(0+) \ne DP_\mu(1-)$. 

From the point of view of the general framework, with gaps, this
corresponds to only one gap, between $1$ and $0$. The inside momentum
loops will also have one gap each, between {\em their} initial and
finite points respectively $S,0$ and $1,S$. In other words, the "loop"
has a topology of an interval, rather than a circle. Cutting the
interval at some inside point we get two intervals, so that the 
topology is trivially preserved.   

The Fourier transformation is defined as follows
\be
\SM{P,\psi,\psi'}_{A}^{B} = \int {\cal D} X \EXP{\i \int_A^B X_\mu(S) d 
P_\mu(S)} \SW{X},
\ee
where $\psi, \psi'$ are boundary values of the Fermi part of the path 
$X_\mu(S)$.

After the Fourier transformation the loop equation reads
\be
 DP_\mu(0) \SM{P, \psi,\psi'}_0^1 \,\int_0^1 d P_\mu(S)   = 
 -N_c g_0^2 \,\int_0^1 d S 
\int d^d \psi(s) \fbyf{\SM{P,\psi,\psi(s)}_0^S 
\SM{P,\psi(s),\psi'}_S^1}{P_\mu(0)\delta P_\mu(S)}.
\ee
This equation is not as simple as it looks. Several things needs too 
clarified.

The coordinate  $\delta$ function disappeared in the same way as in
the ordinary momentum loop equation \ct{Mig86}. The product $
\delta^d(X(0)) \delta^d(X(S)-X(0)) $ can be replaced by $ \delta^d(X(0))
\delta^d(X(S)) $ which is just what is needed to get the correct
integrand in the product of the two Fourier transforms. The $
\delta^d(X(0)) $ function eliminates translations in $\SW{X}_0^S$ and
the other one, $  \delta^d(X(S)) $ does the same for $\SM{X}_S^1$. 

Note that in the last delta function the $\theta $ term in superfield
is present, which adds the terms with gradients of the delta
function. This is not a problem, since the same SUSY $\delta$ function
was included in the definition of the Fourier transform. We had to do
it to preserve the local SUSY. By virtue of SUSY the choice of the
point to fix on a loop is arbitrary, so that we could replace $
\delta^d(X(S)) $ by $\delta^d(X(1)) $in the Fourier integral for
$\SM{P}_S^1$. 

Some subtleties of the factorization of the Fourier integrals are 
discussed in Appendix D.

Note, that there was no delta function for the Fermi part of the 
loop, therefore the integration over $\psi(s)$ was not eliminated. There is 
no explicit dependence of these values in our equation, just the 
integration. Therefore, we can look for the diagonal solution
\be
\SM{P,\psi,\psi'}_A^B = \delta^d\lrb{\oh(\psi + \psi')}\SM{P+}_A^B + 
\delta^d\lrb{\oh(\psi - \psi')}\SM{P-}_A^B.
\ee
Here $\SM{P\pm}$ correspond to the two types of boundary conditions
\be
\SM{P\pm}_A^B = \int d^d \psi \SM{P,\psi,\pm \psi}_A^B.
\ee
We used the fact that $\delta^d(0) = 0$ for the Fermi variables.

Thus $\SM{P+}$ describes the chiral anomaly, and $\SM{P-}$ - the 
vacuum energy (in the latter case one should also include the quark 
kinetic energy -- see below).

Then, the $\psi$ delta functions integrate out, and we are left with the 
single equation
\be
 DP_\mu(0) \SC{P}_0^1 \,\int_0^1 d P_\mu(S)   = 
 2 N_c g_0^2 \,\int_0^1 d S  \fbyf{\SC{P}_0^S 
\SC{P}_S^1}{P_\mu(0)\delta P_\mu(S)},
\ee
where we  introduced the chiral functional
\be
\SC{P} = \oh \lrb{\SM{P-} + \SM{P+}}
\ee
which corresponds to projection operator $\oh\lrb{1 + \gamma_5}$
inserted into the Dirac matrix trace. This projection operator
commutes with the Lorentz generators
\be
\sigma_{\mu\nu} = \inv{4\i}\COM{\gamma_\mu}{\gamma_\nu},
\ee
which enter the Pauli term $\sigma_{\mu\nu} F_{\mu\nu}$ in the square
of the Dirac operator. Hence, the chiral functional corresponds to the
chiral  generators $\oh\lrb{1 + \gamma_5}\sigma_{\mu\nu}$.
The usual even/odd functionals correspond to the even/odd parity part
of the chiral functional.

In spite of chiral loop equation being parity even, we have to include the
terms $  \E{\mu} $ in the coefficients. These
terms are fixed by the chiral anomaly, as discussed above.
We shall come back to this issue below, after we introduce external field.

The SMLoop equation can be simplified if we introduce the SUSY area 
derivative
\be
\fbyf{\SC{P}_A^B}{P_\mu(S)} = D P_\nu(S) 
\pbyp{\SC{P}_A^B}{\Sigma_{\mu\nu}(S)},
\ee
or, in terms of differential forms
\be
\pbyp{\SC{P}_A^B}{P_\mu(S)} = d P_\nu(S) 
\pbyp{\SC{P}_A^B}{\Sigma_{\mu\nu}(S)},
\ee

Later we compute the area derivatives for the terms of the SUSY Taylor
expansion of the SMLoop. As discussed before, we assume the smooth
non-periodic function $P(U)$, so that $P_\mu(S+) = P_\mu(S-)$ but
$P_\mu(1) \ne P_\mu(0)$. 

In the SMLoop equation  both linear and bilinear terms involve the $ d
P_\mu(0) = \oh d P_\mu(1-) + \oh d P_\mu(0+)$ form. {\em Both } of
these forms being arbitrary, we could "cancel" them, i.e. leave only
the tensor coefficients multiplying each form. 

This yields {\em two}  equations 
\be
\label{VSL}
0 = \int_0^1 d  P_\mu(S) \lrb{ \delta_{\lambda\mu} \SC{P}_0^1 + N_c 
g_0^2\, \pbyp{\SC{P}_0^S \SC{P}_S^1}{\Sigma_{\mu\nu}(\xi)\d 
\Sigma_{\lambda\nu}(S)}}, 
\ee
where $ \xi = 0$ or $\xi = 1$.

\section{The External Field}

The problem with the momentum loop dynamics is that it is non-perturbative.  
The zeroth approximation of perturbation theory corresponds to singular 
momentum loop which singularity we expect to disappear in the full solution.  
The only way we can study this full solution so far is to expand in powers 
of momenta.  The expansion coefficients are expected to be proportional to 
the powers of confinement radius which is infinite in perturbation theory.

The loop equation being nonlinear, we can never know whether the 
expansion we build, really matches the perturbative QCD as it should at 
confinement scale of momenta. What if this is a wrong solution, some 
strong coupling artifact, like the ones in lattice gauge theory?

There is a following way around this obstacle. Add the external 
Abelian constant field $B_{\mu\nu}$, then at large $B_{\mu\nu}$ the 
asymptotic freedom will hold, and we will know the solution. At any 
finite field the SMLoop will be still expandable in powers of momenta.

The $B$- dependence of expansion coefficients can be found from the 
following identities
\be
\label{DDB}
\pbyp{\SC{P}_A^B}{B_{\mu\nu}} = \ih \int_A^B dS_1 d S_2 \Theta(S_1,S_2) 
\fbyf{\SC{P}_A^B}{P_\mu(S_1)\delta P_\nu(S_2)}.
\ee
These identities follow directly from the original functional integral, 
with the Abelian external field factor inserted
\be
\TEXP{\i \int_A^B d X_\mu(S) B_\mu(X(S))} = \EXP{\ih B_{\mu\nu}\int_A^B 
\Theta(S_1,S_2) dX_\mu(S_1) dX_\nu(S_2)}.
\ee
In the Fourier integral one could replace $dX_\mu(S) $ by $\i 
\ff{P_\mu(S)}$.

It is convenient to introduce the area derivatives, then one finds in 
\rf{DDB} after some algebra
\bea
\pbyp{\SC{P}_A^B}{B_{\mu\nu}} = \frac{\i}{4} \lrb{\pp{\Sigma_{\mu\nu}(A)} 
+ \pp{\Sigma_{\mu\nu}(B)}} \SC{P}_A^B + \br
\ih \int_A^B d P_\lambda(S_1) d P_\delta(S_2) \Theta(S_1,S_2) 
\pbyp{\SC{P}_A^B}{\Sigma_{\mu\lambda}(S_1)\d \Sigma_{\nu\delta}(S_2)}.
\eea

As for the loop equations, they are modified in an obvious way, by adding 
the derivatives  of this exponential to the operators $\ff{X_\mu(S)}$.

The resulting equations read
\bea
0 =\int_0^1 d  P_\mu(S) \lrb{  \lrb{\delta_{\rho\lambda} + \i 
B_{\rho\sigma}\pp{\Sigma_{\sigma\lambda}}} \lrb{\delta_{\rho\mu} + \i 
B_{\rho\delta}\pp{\Sigma_{\delta\mu}}} \SC{P}_0^1 + \brr
+  N_c g_0^2\,  \pbyp{\SC{P}_0^S \SC{P}_S^1}{\Sigma_{\mu\nu}(\xi)\d 
\Sigma_{\lambda\nu}(S)}}, 
\eea
where $ \xi = 0$ or $\xi = 1$.

It is convenient to pull out of the SMLoop the exponential, corresponding 
to the free quark theory
\be
\SC{P}_A^B \Ra \EXP{\ih B^{-1}_{\mu\nu} \int_A^B P_\mu(S) d P_\nu(S)} 
\SC{P}_A^B.
\ee
This factor being multiplicative around the loop, it will cancel on both 
sides of the loop equation. The only result will be the shift
\be
\label{SH}
\ff{P_\mu(S)} \Ra \ff{P_\mu(S)} + \i B^{-1}_{\mu\nu} D P_\nu(S).
\ee

This shift eliminates the $ DP $ term in the linear part of the momentum 
loop equation, and we get
\bea
B_{\nu\sigma}B_{\nu\delta}\int_0^1 d  P_\mu(S) \pp{\Sigma_{\sigma\lambda}}  
\pp{\Sigma_{\delta\mu}} \SC{P}_0^1 = \br
N_c g_0^2\, \int_0^1 d  P_\mu(S) \lrb{\i B^{-1}_{\mu\nu} + \pp{\Sigma_{\mu\nu}(\xi)}}
\lrb{\i B^{-1}_{\lambda\nu} + \pp{\Sigma_{\lambda\nu}(\xi)}}\SC{P}_0^S 
\SC{P}_S^1, 
\eea
 
Now, the perturbative solution would correspond to iterations of this
equation in $g_0$, starting with the free quark loop. The free quark computations
are presented in Appendix C (see also below). The result for initial
definition of chiral supermomentum loop functional reads
\bea
\SC{P}_A^B \ra  \sqrt{\det \i B} \,\theta\lrb{\sqrt{\det \i
B}} \,\EXP{\ih B^{-1}_{\mu\nu} \int_A^B P_\mu(S) d P_\nu(S)} + \O{g_0^2}.
\eea
It is an interesting challenge to verify how the higher terms of
perturbation theory in a background field are reproduced.

\section{The Position Operator and Nonlinear Symmetry}

Let us proceed with the momentum superloops. The next step in \ct{Mig94}
was the introduction of the position operator $\X{\mu}$ as a connection
in momentum  space.

The corresponding generalization of this Ansatz would be
\be
\label{X}
\SC{P}_A^B = \inv{N_c g_0^2}\,\vac{\TEXP{\i \int_A^B d P_\mu(S) \X{\mu}}}.
\ee

The operator $\X{\mu}$ is an ordinary Bose operator. There are no
restrictions on the commutators $\XX{\mu}{\nu}$, so that these
operators form a free algebra.

This Ansatz involves the superloop in momentum space with constant
non-Abelian connection $\i\X{}$. In components: 
\be
\TEXP{\i \int_0^1 d P_\mu(S) \X{\mu}} =\TEXP{ \int_0^1 d s\lrb{\i \X{\mu} 
p'_\mu -
\oh \XX{\mu}{\nu} \varphi_\mu \varphi_\nu}}. 
\ee
The above definition of the position operator $\X{}$ does not assume
periodicity of the superloop. One could translate $\X{}$ by a constant
vector operator $\hat{C}_\mu$, which commutes with $\X{\mu}$ 
as well as itself:
\bea 
\label{TRANS}
\X{\mu} \Ra \X{\mu} + \hat{C}_\mu, \br
\COM{\hat{C}_\mu}{\hat{C}_\nu} =
\COM{\hat{C}_\mu}{\X{\nu}} = 0.
\eea
As we shall see in a moment, the loop equation stays invariant.

This gauge freedom produces some arbitrary terms in the expectation
values of products of $\X{}$ operators, which, however, should all
cancel out in the momentum superloops and other observables, by virtue
of periodicity of the physical supermomentum. 

Let us now compute the SUSY area derivative in momentum loop space.
The result is a SUSY extension of the Mandelstam formula for the gauge 
potential $\i \X{\mu}$
\bea
N_c g_0^2\,\pbyp{\SC{P}_A^B }{\Sigma_{\mu\nu}(S)} =\br -\vac{\TEXP{\i 
\int_A^S  d P_\mu(U) \X{\mu} } \XX{\mu}{\nu}\TEXP{\i \int_S^B  d P_\mu(V) 
\X{\mu}}}. 
\eea
In components it follows from the direct variation. In a superfield
language it is even simpler. Applying the $\pp{P_\mu(S)}$ operator we
get sum of terms with 
\be
\X{\mu}\int_{S_{l-1}}^{S_{l+1}}  d   \delta(S_l -S)  =
\X{\mu}\lrb{\delta(S_{l+1}-S) - \delta(S_{l-1}-S)} . 
\ee
The first term eliminates the integration over $S_{l+1}$, which yields
$ D P_\nu(S) \X{\mu}\X{\nu} $, the second one yields  $ -D P_\nu(S)
\X{\nu}\X{\mu} $. The rest of terms add up to the product of two
ordered exponentials. 

Now we are in a position to study the superloop equations \rf{VSL}. 
The $\xi =0$ equation becomes
\bea
\int_0^1 d P_\mu(S) \lrb{  \delta_{\mu\lambda} \vac{\TEXP{\i\int_0^1 d 
P_\mu(U) \X{\mu}}} + \brr
\vac{\TEXP{\i\int_0^S d P_\mu(V) \X{\mu}}\XX{\mu}{\nu}} \brr
\vac{\XX{\lambda}{\nu}
\TEXP{\i\int_S^1 d P_\mu(W) \X{\mu}}} +\brr
\vac{\TEXP{\i\int_0^S d P_\mu(V) \X{\mu}}
\XX{\mu}{\nu}\XX{\lambda}{\nu}}\brr
\vac{\TEXP{\i\int_S^1 d P_\mu(W) \X{\mu}}}} = 0.
\eea

Using the factorization of the ordered exponent we can write it as
\be
\vac{\int_0^1 d P_\mu(S) \TEXP{\i\int_0^S d P_\mu(V) \X{\mu}}
\hat{H}(\X{})_{\mu\lambda} \TEXP{\i\int_S^1 d P_\mu(W) \X{\mu}}} = 0, 
\ee
where
\be
\label{HDEF}
\hat{H}(\X{})_{\mu\lambda} =  \delta_{\mu\lambda} + \XX{\mu}{\nu}
\AC{\P}{\XX{\lambda}{\nu}}, 
\ee
is some operator, related to $\X{}$.

But this is the symmetry property! The transformation
\be
\delta \X{\mu} = \delta\epsilon_\lambda \hat{H}(\X{})_{\mu\lambda},
\ee
will result in the same variation of our Ansatz. Therefore, we
conclude that the momentum loops $\SC{P}$ are invariant with respect
to these transformations with arbitrary infinitesimal vector
$\delta\epsilon_\mu$. We could consider the family of operators
$\X{\mu}(t)$ parameterized  by arbitrary path $\epsilon(t)$ and
obeying the equations 
\be
\d_t \X{\mu} = \d_t \epsilon_\lambda \,\hat{H}(\X{})_{\mu\lambda}.
\ee
The correlation functions must be independent of $\epsilon(t)$. The derivative
\bea
N_c g_0^2 \,\pbyp{\SC{P}_0^1}{t} = \br
\i \int_0^1 d P_\mu(S) \vac{
\TEXP{\i\int_0^S d P_\mu(V) \X{\mu}} \hat{H}(\X{})_{\mu\lambda} \d_t
\epsilon_\lambda \TEXP{\i\int_S^1 d P_\mu(W) \X{\mu}}} = 0, 
\eea
for arbitrary non-periodic function $P_\mu(S)$.

In the same way, the $\xi=1$ equations are equivalent to the symmetry 
transformations with
\be
\label{HPDEF}
\hat{H}'(\X{})_{\mu\lambda} = \delta_{\mu\lambda} +
\AC{{\P}}{\XX{\lambda}{\nu}}\XX{\mu}{\nu} , 
\ee

The difference of these transformations is the commutator
\be
\hat{H}(\X{})_{\mu\lambda}-\hat{H}'(\X{})_{\mu\lambda} =
\COM{\XX{\mu}{\nu}}{\AC{{\P}}{\XX{\lambda}{\nu}}}. 
\ee

What is the meaning of these observations? The loop equations were
derived as Schwinger-Dyson equations for the large $N_c$ Yang-Mills
theory. Apparently, this theory is equivalent to the Quantum Mechanics
of the operators $\X{}$, with some hidden nonlinear symmetry. This is
not a symmetry of the loop equations, this symmetry is {\em all} these
equations are about.  

Any representation of this symmetry solves the loop
equations. We could introduce the generators $\hat{\Gamma}_\mu, \hat{\Gamma}'_\mu$ such
that
\bea
\label{SYM}
\COM{\X{\mu}}{\hat{\Gamma}_\lambda} =  \hat{H}(\X{})_{\mu\lambda},\br
\COM{\X{\mu}}{\hat{\Gamma}'_\lambda} =
\hat{H}'(\X{})_{\mu\lambda},\br
\hat{\Gamma}_\lambda \r0 =\hat{\Gamma}'_\lambda \r0 = 0, \br
\l0 \hat{\Gamma}_\lambda = \l0 \hat{\Gamma}'_\lambda = 0.
\eea

The Fock space is defined as the set of words
\bea
\left |\mu_1\dots\mu_m \right \rangle = \ad{\mu_1}\dots\ad{\mu_m}\r0,\br
\left \langle \nu_n\dots\nu_1 \right| = \l0 a_{\nu_n}\dots a_{\nu_1},\br
\left \langle \nu_n\dots\nu_1 \right| \left |\mu_1\dots\mu_m \right\rangle = \delta_{mn}\,\prod_i^n
\delta_{\mu_i\nu_i}.
\eea
The position operator according to Voiculesku can be chosen to be
\be
\X{\mu} = a_\mu + \sum_{k=1}^\8 Q_{\mu,\mu_1\dots\mu_k} \ad{\mu_1}\dots \ad{\mu_k},
\ee
where the first term reduces the word by one letter, and the
second term expands it by arbitrary number of letters. That is
\bea
\left \langle \nu_n\dots\nu_1 \right|\X{\mu} \left |\mu\nu_1\dots\nu_{n}\right \rangle
=1,\br
\left\langle \nu_n\dots\nu_1 \right|\X{\mu}\left |\nu_k\dots\nu_{n}\right\rangle
= Q_{\mu,\nu_1\dots\nu_{k-1}},
\eea
the rest of matrix elements vanishing.
As for the generators, those have arbitrary matrix elements, between
any finite words.

Let us write down  explicit set of relations between matrix elements.
First, consider the VEV of these equations, we get
\be
\label{VAC}
\vac{\XX{\mu}{\nu}\XX{\nu}{\lambda}} =  \delta_{\mu\lambda}. 
\ee
We used the fact that $\vac{\XX{\mu}{\nu}} = 0 $, which follows from
space symmetry in more than two dimensions (in two dimensions this
could be proportional to $\epsilon_{\mu\nu}$).

The next equation follows when the matrix elements between vacuum and
arbitrary word are taken. We get
\be
\l0 \X{\mu} \hat{\Gamma}_\lambda \left| \mu_1\dots\mu_n \right\rangle
= 0.
\ee
All the creation terms in $\X{\mu}$ drop here, so we simply find
\be
\left \langle \mu \right | \hat{\Gamma}_\lambda = 0.
\ee
In other words, not only vacuum state is annihilated by the generator
$ \hat{\Gamma}_\lambda $, any (left) letter is annihilated as
well. For the second generator $ \hat{\Gamma}'_\lambda $ we get the
similar equation
\be
\hat{\Gamma}'_\lambda \X{\mu} \r0 = 0.
\ee
However, the operator $\X{\mu}$ generates the tower of word states
from the right vacuum, so that this equation is not so simple.

Finally, the non-vacuum matrix elements of the symmetry relations read
\bea
\left\langle W\left| \COM{\X{\mu}}{\hat{\Gamma}_\lambda} \right | W'\right\rangle  = 
 \delta_{\mu\lambda}\delta_{W,W'} +
\left\langle W\left|
\XX{\mu}{\nu} \right|0\right\rangle \,
\left\langle \COM{\lambda}{\nu}\left|\right | W'\right\rangle,\br
\left\langle W\left| \COM{\X{\mu}}{\hat{\Gamma}'_\lambda} \right | W'\right\rangle  = 
 \delta_{\mu\lambda}\delta_{W,W'} +  \left\langle W\left|
\XX{\lambda}{\nu} \right|0\right\rangle \,
\left\langle \COM{\mu}{\nu}\left|  \right | W'\right\rangle,
\eea
with obvious notation
\be
\left\langle \COM{\mu}{\nu}\right| \equiv \left\langle \mu\nu\right| -
\left\langle \nu\mu\right|.
\ee
Note that these equations are only quadratic in unknown matrix
elements of $\X{}$. All the higher order terms canceled. Note also
that there is an infinite set of bilinear equations
\bea
\left\langle W\left| \COM{\X{\mu}}{\hat{\Gamma}_\lambda} \right | W'\right\rangle  = 
\left\langle W\left| \COM{\X{\mu}}{\hat{\Gamma}'_\lambda} \right |
W'\right\rangle  =  \delta_{\mu\lambda}\delta_{W,W'},
\eea
for all the words $W'$ with $ n \neq 2$ letters. The truly nonlinear
equations are those with $ n = 2$
\bea
\left\langle W\left| \COM{\X{\mu}}{\hat{\Gamma}_\lambda} \right|\alpha\beta\right\rangle  =
 \delta_{\mu\lambda}\left\langle W\left| \right
|\alpha\beta\right\rangle  + \left\langle W\left|
\COM{\X{\mu}}{\X{\beta}} \delta_{\lambda\alpha} -
\COM{\X{\mu}}{\X{\alpha}} \delta_{\lambda\beta} \right|0\right\rangle,\br
\left\langle W\left| \COM{\X{\mu}}{\hat{\Gamma}'_\lambda} \right|\alpha\beta\right\rangle  = 
 \delta_{\mu\lambda}\left\langle W\left| \right
|\alpha\beta\right\rangle  + \left\langle W\left|
\COM{\X{\lambda}}{\X{\beta}} \delta_{\mu\alpha} -
\COM{\X{\lambda}}{\X{\alpha}} \delta_{\mu\beta} \right|0\right\rangle.
\eea

At present we do not know how to solve these equations in closed
form. However, the recurrent equations for the correlation functions
of products of $\X{}$ can be derived and solved one after another. 

\section{Operator Expansion}

These recurrent equations follow directly from the symmetry property
\be
\sum_{l=1}^n \vac{\X{\mu_1}\dots \hat{H}(\X{})_{\mu_l\lambda}\dots 
\X{\mu_n}} = 0.
\ee
At $ n = 1$ we have the previous equation \rf{VAC}.
In the higher orders we get similar recurrent equations.
The highest rank tensor here come from the term with $ l=n$. Moving
all the rest to the \RHS and skipping the terms which vanish
identically, and using the $n=1$ equation we get a recurrent equation

\bea
\vac{}\vac{\X{\mu_1}\dots \X{\mu_{n-1}} \XX{\mu_n}{\nu}\XX{\lambda}{\nu}} 
=\br
- \sum_{l=2}^{n} \delta_{\mu_l\lambda}\vac{\X{\mu_1}\dots
 \X{\mu_{l-1}}\X{\mu_{l+1}}\dots \X{\mu_n}}\br
- \sum_{l=3}^{n-2} \vac{\X{\mu_1}\dots \XX{\mu_l}{\nu}} 
\vac{\XX{\lambda}{\nu}\dots \X{\mu_n}}\br
- \sum_{l=3}^{n-2}\vac{\X{\mu_1}\dots \XX{\mu_l}{\nu}\XX{\lambda}{\nu}} 
\vac{\dots \X{\mu_n}}.
\eea

The second transformation leads to the double commutator equation
\be
\sum_{l=1}^n 
\vac{\X{\mu_1}\dots\COM{\XX{\mu_l}{\nu}}{\AC{\P}{\XX{\lambda}{\nu}}}\dots 
\X{\mu_n}} = 0. 
\ee
There are two highest rank tensor terms, which cancel in virtue of
cyclic symmetry. Therefore, the second equation reads 
\be
\sum_{l=3}^{n-2} 
\vac{\X{\mu_1}\dots\COM{\XX{\mu_l}{\nu}}{\AC{\P}{\XX{\lambda}{\nu}}}\dots 
\X{\mu_n}} = 0 
\ee

The first terms of this expansion are easy to find. The most general
solution for the four point function reads 
\be
\vac{\X{\mu_1}\X{\mu_2}\X{\mu_3}\X{\mu_4}}  =
-\frac{1}{2(d-1)}\,\delta_{\mu_1\mu_3}\delta_{\mu_2\mu_4} + symmetric.
\ee
The symmetric terms, proportional to
\be
\delta_{\mu_1\mu_2}\delta_{\mu_3\mu_4} + 
\delta_{\mu_1\mu_3}\delta_{\mu_2\mu_4} + 
\delta_{\mu_1\mu_4}\delta_{\mu_2\mu_3},
\ee
are left ambiguous. These are the gauge terms, which cancel in
observables by virtue of translational symmetry \rf{TRANS}. The
correlation functions of Abelian operators $\VEV{C_{\mu_1}\dots
C_{\mu_n}}$ produce symmetric terms. 

In four dimensions there is an extra term
\be
N_c g_0^2 \epsilon_{\mu_1\mu_2\mu_3\mu_4}
\ee
in the four point function, coming from chiral anomaly. In two
dimension the similar term
\be
N_c g_0^2 \epsilon_{\mu_1\mu_2}
\ee
would contribute to the two point function.

The anomalous terms have perturbative normalizations, not affected by effects of
strong interactions. The normalization can be recovered from
\rf{ANOM}, keeping in mind the extra factor of $\inv{N_c g_0^2}$ in
definition of the vacuum average \rf{X}.

The two-point correlation in four dimensions remains undetermined. This introduces
arbitrary parameter  of dimension of square of length 
\be
\vac{\X{\mu}\X{\nu}} = A \vac{}\delta_{\mu\nu}.
\ee

In the $n=3$ order we get the  equation for the $6$-th rank tensor
\be
\vac{\X{\mu_1}\X{\mu_{2}}\XX{\mu_3}{\nu}\XX{\lambda}{\nu}} = -A
\lrb{\delta_{\mu_1\mu_2}\delta_{\mu_3\lambda} +
\delta_{\mu_1\mu_3}\delta_{\mu_2\lambda}}. 
\ee

The computations of these tensors was done by means of the \MAT
algorithms of the previous work (see files attached to the hep-th
source  of \ct{Mig94}).
Again, there are some free coefficients, which (partly) drop in
observables.     

The non-commutative probability theory goes as usual with this
definition of $\X{}$ correlators. The operator expansion  goes  as
above, with coefficients $Q_{\mu_n,\dots\mu_1}$ given by
planar connected moments  $\vac{\X{\mu_1}\dots\X{\mu_n}}_{pl. conn} $.

\section{Glueball Equation}

As it was pointed out in \ct{Mig94} the equation for the glueball
excitation wave-function $\Psi[C]$ corresponds to linearization of the
loop equation, which can be achieved by variations of the position
operator. Let us elaborate this idea in the context of the
supermomentum loop equation.

First, let us identify the total momentum dependence of the glueball
loop  equation. The coordinate superloop equation had the form of
\be
\hat{L}_C\lrb{\SW{.}} = \hat{R}_C\lrb{\SW{.},\SW{.}},
\ee
where $\hat{L} $ is a linear operator, and $\hat{R}$ is a bilinear
one.

The initial form of the $\hat{L}$ operator was
\be
\hat{L}_C\lrb{\SW{.}} =
\int_{-0}^{+0} d S \fbyf{ \SW{X}_0^1}{X_\mu(0)\delta X_\mu(S)},
\ee
which involved integration over infinitesimal interval $\lrb{-0, +0}$
around the point $0$ where the first variation took place. This
operator is local, it picks only the delta terms, related to the Y-M
equations of motion.

Now, in virtue of translation invariance
\be
\oint d S \fbyf{\SW{C}}{X_\mu(S)} = 0,
\ee
which allowed us to replace the local operator by a nonlocal one
\be
\hat{L}_C\lrb{\SW{.}} =
-\int_{0}^{1} d S \fbyf{ \SW{X}_0^1}{X_\mu(0)\delta X_\mu(S)},
\ee
with integration covering the rest of the loop.
This was the operator we used above, in transforming superloop equation
to the momentum space.

When the glueball wave function is studied, the translation invariance
is replaced by fixed momentum condition
\be
\oint d S \fbyf{\Psi[C]}{X_\mu(S)} = \i k_\mu\Psi[C].
\ee
This will result in extra term in the momentum loop operator
\be
\hat{L}_P\lrb{\Psi} = D P_\mu(0) \,\lrb{-k_\mu +\int_0^1 d
P_\mu(S)}\Psi[P].
\ee

Repeating all the previous steps for the linearized momentum loop
equation, we find
\bea
\label{PSIEQ}
k_\lambda \Psi[P]_0^1  = 
\int_0^1 d  P_\mu(S) \lrb{ \delta_{\lambda\mu} \Psi[P]_0^1 + N_c 
g_0^2\, \pbyp{\lrb{\SC{P}_0^S \Psi[P]_S^1 + \Psi[P]_0^S \SC{P}_S^1}}
{\Sigma_{\mu\nu}(\xi)\d \Sigma_{\lambda\nu}(S)}}. 
\eea

Let us prove that the following Ansatz solves this equation
\be
\Psi[P]_A^B = \int_A^B d P_\mu(S) \vac{\TEXP{\i\int_A^S \X{\nu}d P_\nu(V)}
\Y{\mu}\,\TEXP{\i\int_S^B  \X{\lambda}d P_\lambda(W)}},
\ee
provided the operator $\hat{Y}_\lambda$ satisfies the commutation relations
\be
\label{GLU}
\COM{\hat{\Gamma}_\lambda}{\Y{\mu}}
=\COM{\hat{\Gamma}'_\lambda}{\Y{\mu}} = k_\lambda \Y{\mu}. 
\ee
The idea is to take a variation $\delta \SC{P} = \Psi[P]$ of initial
operator representation \rf{X} with $\delta \X{\mu} = N_c g_0^2
\hat{Y}_\mu$, to reproduce the right side of the glueball loop
equation \rf{PSIEQ}. The left side comes about from the commutator \rf{GLU}.

The formal proof follows from the identity
\be
0 \equiv \int_0^1 d P_\mu(S) \vac{\COM{\TEXP{\i\int_0^S \X{\nu}d P_\nu(V)}
\Y{\mu}\,\TEXP{\i\int_S^1  \X{\lambda}d P_\lambda(W)}}{\hat{\Gamma}_\lambda}}
\ee
and the same for $\hat{\Gamma}'_\lambda$. Commuting the  generators with
ordered exponentials as follows
\bea
\COM{\TEXP{\i\int_A^B \X{\nu}d P_\nu(V)}}{\hat{\Gamma}_\lambda} =\br
\i \int_A^S d P_\mu(U)\TEXP{\i\int_A^U \X{\alpha}d P_\alpha(W)}
\hat{H}_{\mu\lambda}(\X{})\,\TEXP{\i\int_U^B \X{\beta}d P_\beta(T)}.
\eea
and using the definitions \rf{HDEF},\rf{HPDEF}, \rf{GLU} and algebra of
differential forms \rf{COMP},\rf{MULT}, it  is straightforward to verify
the glueball loop equation.

The linear equations \rf{GLU} represent the eigenvalue equations for
the glueball spectrum in this theory. The states $\hat{Y}_\mu\r0$
represent the momentum eigenstates,
\bea
\hat{\Gamma}_\alpha
\hat{Y}_\mu\r0 = k_\alpha \hat{Y}_\mu\r0,\br \hat{\Gamma}'_\alpha
\hat{Y}_\mu\r0 = k_\alpha \hat{Y}_\mu\r0.
\eea

We see, that operators $\hat{\Gamma}_\alpha,\hat{\Gamma}'_\alpha$ play the role of
spatial translation generators of the endpoint of ``QCD string''. They have to be
found from above nonlinear operator algebra. In general, they do not
commute, so these are not the Poincare translation generators.

Clearly, the commutators must annihilate these states
\bea
\COM{\hat{\Gamma}_\alpha}{\hat{\Gamma}_\beta} \hat{Y}_\mu\r0 =0,\br
\COM{\hat{\Gamma}'_\alpha}{\hat{\Gamma}'_\beta} \hat{Y}_\mu\r0 =0,\br
\COM{\hat{\Gamma}_\alpha}{\hat{\Gamma}'_\beta} \hat{Y}_\mu\r0 =0.
\eea
which yields restrictions on the Voiculesku coefficients of
$\hat{Y}_\mu$ operators.

\section{Quark Propagating Around the Superloop}

Let us now have a look at the negative parity current correlators,
which up to  normalization and standard Chan-Paton factor $ \tr
\tau_{a_1} \dots \tau_{a_n} $ are given by the following integral 
\bea
\Gamma_n^{\nu_1\dots\nu_n}\lrb{k_1,\dots k_n} = \br
\i^n  \int_0^1  
\pp{K_{\nu_1}(S_1)}\dots  \pp{K_{\nu_n}(S_n)}
\vac{\TEXP{\i \int_0^1 d  K_\mu(S) \X{\mu} }},
\eea
where it is implied that after taking all functional derivatives the 
supermomentum
\be
K_\mu(S) =\sum_{i=1}^n k_\mu^{i} \Theta(S_i,S).
\ee
So, the value of the supermomentum also depends upon $S_1,\dots
S_n$.  

There is a systematic way of handling such superpath the ordered
exponentials, suggested in \ct{Tseytlin}. In our context this looks as
follows. 
We introduce the Quark superfield
\be
Q_W(S) = q_W(s) + \theta \zeta_W(s)
\ee
which is a $\left|ket\right\rangle$  vector in the Fock space of words
$W = \left|\mu_1\dots\mu_n\right\rangle$, and respectively 
\be
\bar{Q}_W(S) = \bar{q}_W(s) + \theta \bar{\zeta}_W(s),
\ee
which is a $\left\langle bra \right|$ vector.
Then we represent the SMLoop as follows
\bea
\vac{\TEXP{\i \int_0^1 d K_\mu(U) U \X{\mu} }} = \br
\int {\cal D} \bar{Q} {\cal D} Q \EXP{\int_0^1  \bar{Q}(U) \lrb{ d
Q(U) -\i  d K_\mu(U) \X{\mu} Q(U)}} \vac{Q(0) \otimes \bar{Q}(1)}. 
\eea
The proof essentially repeats the arguments of  \ct{Tseytlin}, with
some improvements and corrections. For brevity we will skip the
indexes $W$, so that, e.g.  the propagator 
\be
G(S,S') = \VEV{Q(S) \otimes \bar{Q}(S')}_{\bar{Q},Q},
\ee
represents the operator in Fock space with matrix elements
\be
\VEV{W\right|G(S,S')\left|W'} = 
\VEV{\mu_1\dots\mu_n\right|G(S,S')\left|\mu'_1\dots\mu'_{n'}}.
\ee

The bare propagator $G^0(S,S')$ of this Quark satisfies an equation
\be
D G^0(S,S') = \delta(S'-S) ,
\ee
which we know how to solve
\be
G^0(S,S') = \Theta(S,S').
\ee

The higher vacuum loops all vanish by virtue of the identities
\bea
\Theta(S,S')\Theta(S',S) = 0,\br
\Theta(S,S')\Theta(S',S'')\Theta(S'',S) = 0,\br
\dots
\eea
This leaves only one vacuum loop, corresponding to first order diagram
\be
\int_0^1 d  K_\mu(S)\Theta(S,S)  \tr X_\mu = \oh\int_0^1 d  K_\mu(S) \tr 
X_\mu
\ee
which vanishes  by virtue of periodicity of $K_\mu(S)$, (plus the
trace vanishes due to Lorentz symmetry). 

The complete Greens function is a solution of the equation
\be
\label{DGEXP}
D G(S,S') = \delta(S'-S) - \i (D K_\mu(S)) \X{\mu} G(S,S').
\ee
The formal solution of this equation is given by the the ordered 
exponential
\be
G(S,S')  = \TEXP{\i \int_{S}^{S'} d K_\mu(U) \X{\mu} },
\ee
which is just what we need.

The variations with respect to supermomentum now act on usual rather
than the ordered exponential with the result 
\be
\i \pp{K_\mu(S)} \Ra d \lrb{\bar{Q}(S) X_\mu Q(S)} = d \bar{Q}(S)X_\mu 
Q(S) + \bar{Q}(S)X_\mu d Q(S),
\ee
which plays the role of the vector current of our superloop theory.

The multicurrent amplitude reads
\bea
\Gamma_n^{\nu_1\dots\nu_n}\lrb{k_1,\dots k_n} = \br
 \int_0^1    \VEV{d^n \lrb{\bar{Q}(S) X_{\lbrace \nu\rbrace} Q(S)} 
\vac{Q(0) \otimes \bar{Q}(1)}}_{\bar{Q},Q}.
\eea
One may try to integrate  by parts here, replacing the covariant
derivatives of Quark currents by those acting on the $\Theta$
functions in our the ordered integral. However, there is some more
$S_i$ dependence, which comes from external supermomentum
$K_\mu(S)$. This brings down more Quark currents $ D\bar{Q}\X{\mu}Q$,
which does not simplify formulas. 

The Gaussian average over $\bar{Q},Q$ reduces to super-graphs with
propagator $G(S,S')$ in external field $  \X{\mu} D K_\mu(S)$. Now,
once all variations are done, we can set the supermomentum to its
value, after which this external field reduces to the sum of $\delta$
functions in superspace.

After this the ordered exponential of the integral of sum of the $\delta$ function
terms will reduce to the product of the vertex operators $ 
\prod_l \EXP{\i k_\mu^l\bar{Q}(S_l) \X{\mu}Q(S_l)}$.

The flavor vector current correlation functions are given by the
multiple superspace integrals of the $\bar{Q},Q$ average. This average
reduces to sum of the {\em connected graphs}. All the vacuum graphs
vanish in our 1D supergravity, as discussed above. 

The generic connected graph looks the same as the string theory
graphs, with currents represented by
\be
\label{CUR}
J_\mu^a(k,S) \ra \tau^a \otimes d\lrb{\bar{Q}(S) X_{\mu} Q(S)}
\EXP{\i k_\mu\bar{Q}(S) \X{\mu}Q(S)},
\ee
and propagators $\Theta(S,S')$.

\section{SUSY and Chiral Symmetry}

In the vector sector (anti-periodic boundary conditions) we have to
keep the kinetic term in the loop Lagrangian. 

The general formula \rf{GENERAL} for the current correlations involves
an extra operator $Z$, which we replaced by $1$ in the topological
sector. Using the above quark superfield theory we could replace this
operator by 
\bea
Z = \lim_{\alpha \ra 0}\int_0^\8 d T T^{\alpha-1}\lrb{1 + \alpha \ln
T}\EXP{-T m_0^2 - \frac{\hat{L}}{T}},\br 
\hat{L} = \frac{1}{4}\int_0^1 d S D \lrb{\bar{Q}(S) \hat{X}_\mu Q(S)}
D^2  \lrb{\bar{Q}(S) \hat{X}_\mu Q(S)}, 
\eea
which will result in extra quartic interaction.

What could be an effect of this interaction? Formally,  one can
develop the perturbation expansion  and compute all the arising $T$
integrals 
\bea
\int_0^\8 d T T^{\alpha-n-1}\lrb{1 + \alpha \ln T}\EXP{-T m_0^2} \ra 
\frac{(-m_0^2)^n}{n!} \lrb{\frac{\psi(n+1)}{n!} - \ln m_0^2}.
\eea
We see, that these integrals contain the increasing powers of the bare 
mass.

However, the remaining superspace integrals may diverge, because the
local SUSY is broken down to the global one. Specifically,  the
singularities at small intervals $s-s' \ra 0$ now do not match the
number of integrations. Extra operator $D^2 = \pp{s}$ produces
ultraviolet divergences. Part of these divergences is for real-- one
has to renormalize the mass after all.  

The spontaneous chiral symmetry breaking must come about as a
singularity $\O{\sqrt{m_0^2}}$ in normal channels, but also as extra
poles at zero momenta in the pion channels of vector current
amplitudes we consider.  

This means that the self energy, induced by this quartic interaction,
can remain finite in the zero mass limit, and move the pole to zero
momentum. By power counting it is quite plausible. The bubble diagrams
will contain the singularities, which presumably will be cut at the
new scale, induced by the mass. Solving the Dyson equation for this
new mass scale may lead to the chiral symmetry breaking in the limit
$m_0\ra 0$. 

The divergences produced by the physical mass term are unpleasant. On
the other hand we know that the vector sector in the chiral limit is
finite. The pion poles involve the finite residues which do not need
extra renormalization. Is there a way to avoid these spurious
divergences? 

We suggest the following soft chiral symmetry breaking, without
unnecessary violations of local SUSY. Let us add the constant Abelian
external field with finite density of topological charge. The chiral
symmetry will be  broken by the Adler-Bell-Jackiw anomaly, so that
there will be no massless pion. 

The vacuum energy can be written as the sum of the zero mode
contribution and the finite modes. In the limit of the vanishing quark
mass 
\bea
{\cal E} = -\oh\int_{-\8}^\8 d\lambda \lrb{\nu_0 \delta(\lambda) +
\rho(\lambda)} \ln \lrb{m_0^2 + \lambda^2} \br 
= -\nu_0 \ln m_0 - \oh\int_0^\8 d \lambda  \rho(\lambda)\ln \lrb{m_0^2
+ \lambda^2} \ra -\nu_0 \ln m_0 + const, 
\eea
where $\nu_0 > 0 $ is the net number of zero modes. The leading
divergent term in the chiral limit is the contribution of the zero
modes. Unlike the  chiral anomaly, the net number of zero modes is not
a topological invariant. 

Let us study this number in the free quark theory. The corresponding
Euler-Heisenberg Lagrangian was computed in Appendix B.  
The formula \rf{EH} can be written as
\be
\label{EHCHIRAL}
V \int_0^\8 d T T^{\alpha-1}\lrb{1 + \alpha \ln T}
e^{-T m_0^2}\prod_{i=1}^{\oh d} b_i \coth (2 b_i T),
\ee
where $\pm \i b_i $ are the eigenvalues of $B_{\mu\nu}$. In the chiral
limit the leading logarithmic term comes from the region $b_i^{-1} \ll
T \ll m_0^{-2}$ where one could replace $\coth (2 b_i T) \ra sign
(b_i)$, after which the integral yields 
\be
V \prod |b_i| \int_0^\8 d T T^{\alpha-1}\lrb{1 + \alpha \ln T}
e^{-T m_0^2} \ra -(\gamma + \ln m_0^2) V \left|\sqrt{\det \i B}\right|.
\ee
This is an absolute value of the Pfaffian.

The same result (up  to normalization) could be obtained much simpler,
by neglecting the kinetic energy in the path integral. The path
integral becomes supersymmetric, 
\bea
V \int {\cal D} X \delta^d(X(0))\EXP{\ih B_{\mu\nu}\int_0^1 dX_\mu(S) 
X_\nu(S)} =\br
V \int {\cal D} X \delta^d(X(0))\EXP{ \sum_{i=1}^{\oh d} b_i\int_0^1 
dX_{2i-1}(S) X_{2i}(S)} \propto 
V \prod_{i=1}^{\oh d} |b_i| = V \left|\sqrt{\det \i B}\right|.
\eea
We used the rotational invariance to reduce the antisymmetric
quadratic form to the Jordan form. Then we rescaled half of
components of the superfield $X_{2i}(S) \ra b_i^{-1} X_{2i}(S)$. The
SUSY measure stays invariant, because $STr 1 = 0$, therefore the only
contribution comes from rescaling of the $\delta$ function. 
The proper time integral gives then the logarithmic contribution.

So, we get the absolute value of the topological charge density. The
vacuum  average of this number in QCD does not vanish, unlike the average
topological charge. In the background of external constant flavor
field with nontrivial topology we expect this number to be
proportional to the volume of space, which makes it a leading term in
the chiral limit (it is proportional to $\ln m_0^2$, unlike the
contribution  from the finite modes). 

We could argue that the quark kinetic energy drops in  QCD by the same
reason as in the Euler-Heisenberg Lagrangian, and we arrive at the
locally SUSY theory in the vector sector 
\bea
\Gamma_n^{\nu_1\dots\nu_n}\lrb{k_1,\dots k_n} = \br
-\ln m_0 \int_0^1  \VEV{\vac{Q(0) \otimes \bar{Q}(1)}\,\prod_{i=1}^n 
d\lrb{\bar{Q}(S_i) X_{\nu_i} Q(S_i)} }_{\bar{Q},Q}.
\eea
with the usual supermomentum
\be
K_\mu(U) = \sum_{i=1}^n k_\mu^{i} \Theta(S_i,U).
\ee

The corresponding term in the Quark action  also represents the quartic 
interaction
\be
\ih B_{\mu\nu} \int d S D \lrb{\bar{Q}(S) \hat{X}_\nu Q(S)} 
\lrb{\bar{Q}(S) \hat{X}_\mu Q(S)},
\ee
which preserves the local SUSY.
The only symmetry which gets broken is the Lorentz symmetry, as we now 
have external tensor field. 

One could study the perturbation expansion in  this quartic
interaction and try to identify and sum up the terms responsible for
the spontaneous chiral symmetry breaking in the limit $B_{\mu\nu} \ra
0$. The good thing about this perturbation expansion is that all the
graphs will still be locally SUSY, and therefore calculable. 

This interesting issue calls for a further study.

\section{Summary and Conclusions}.

Let us summarize the status of the superloop theory of the large $N_c$
QCD. Here are the basic features of this theory. 
\begin{itemize}
\item
The chiral momentum loop amplitudes, with projector $\oh\lrb{1 \pm
\gamma_5}$ inside the Dirac trace, satisfy the closed equation,
generalizing the ordinary momentum loop equation, for the case of the
supermomentum loops.
\item
The supermomentum loop equation can be interpreted  as a
nonlinear symmetry relations \rf{SYM} of the position operator $\X{\mu}$ of the
QCD string. This symmetry lead to recurrent relations for the coefficients of the
Taylor expansion of momentum superloop.
\item
The (non-commutative) generators $\hat{\Gamma}_\mu,\hat{\Gamma}'_\mu $ of these symmetry
transformations can be interpreted as string endpoint translation
operators. In particular, the glueball excitations with fixed momentum
$k_\mu$ are related to the eigenstates of these operators with common
vector eigenvalue $k_\mu$.
\item
The position operator can be represented as $ a_\mu $ plus expansion in
$\ad{\mu}$. These operators satisfy Cuntz algebra and annihilate
the vacuum $ \r0 $. Generic state is a word $\prod
a_{\mu_i}^\dagger \r0 $. Expansion coefficients are
given by planar connected moments of $\vac{\prod \X{\mu_i}}$. 
\item
The correlators of the vector quark currents $\VEV{\prod
J^a_{\mu_i}(k^i)}$ in the chiral limit in the background of strong constant
electro-magnetic field are related to (functional derivatives of) the
SMLoop with supermomentum $K_\mu(S) = \sum_i k_\mu^i\Theta(S_i,S)$. 
\item
One could introduce the  Quark superfield at the loop $Q(S)$ in such a
way that the flavor vector current is represented at the loop as
\rf{CUR}.
\item
In the chiral limit the Quark superfield is  Gaussian, with the
propagator, depending on $\X{\mu}$ and momenta of all the
currents. The exact operator solution for this propagator
looks very similar to the good old string diagrams.
\item
The low-energy expansion of flavor current correlators in the chiral 
limit in the background of strong constant electro-magnetic field is also
calculable in terms of these coefficients. There are no singular
integrals involved. All the superloop integrals reduce to rational numbers. These
tedious calculations can be simplified by means of generating function
derived in Appendix E.
\item
In order to account for the chiral symmetry breaking effects in
anomalous channels of these current correlators, one should add the
quark mass $m_0$. This results in extra quartic interaction in our 1D
superfield theory. The interaction constant is proportional to the
$m_0^2$. This interaction breaks the local SUSY down to the global
one, and introduces the extra divergences. Presumably, summing the
leading divergences (bubble graphs?) will account for the chiral
symmetry breaking. 

\end{itemize}

\section{Acknowledgments}

I would like to thank Itzhak Bars, David Gross, Volodya Kazakov, Ivan
Kostov,  Sasha Polyakov, Gabriele Veneziano, Eric Verlinde, Herman 
Verlinde, Shimon Yankelovitch,  and especially Arkady Tseytlin for  
interesting
discussions. Collaboration with Andrey Mikhailov at early stage of
this project was also quite useful. The stimulating atmosphere of CERN
was very important for me, and I am grateful to all participants of
the TH seminar for their interest and support. This work was partially
supported by the National Science Foundation under contract
PHYS-90-21984.

\appendix

\section{One More Representation of a Log}

Let us consider the identity
\be
A^{-\alpha} \Gamma(\alpha) = \int_0^\8 d T T^{\alpha-1} e^{- A T},
\ee
valid for any positive definite operator $A$. Multiplying it by $
\alpha $ ant taking derivative $\pp{\alpha} $ we find 
\be
A^{-\alpha} \Gamma(1 +\alpha) \lrb{\psi(1 + \alpha) - \ln A }=
\int_0^\8 d T T^{\alpha-1} \lrb{1 + \alpha \ln T} e^{- A T}.
\ee
Tending $\alpha \ra +0$,
\be
\gamma + \ln A = -\lim_{\alpha\ra+0} \int_0^\8 d T T^{\alpha-1} \lrb{1 + 
\alpha \ln T} e^{- A T}.
\ee
The formula in the text corresponds to $ A = m_0^2 +
\lrb{\i\gamma_\mu\nabla_\mu}^2 $ which is positive definite regardless
of the possible zero modes of the Dirac operator. 

\section{Euler-Heisenberg Lagrangian}

Let us compute the vacuum energy in constant Abelian field $ B_\nu(X) = 
\ih B_{\mu\nu} X_\mu$.

The quadratic forms in exponential of \rf{VACEN} are the same for Bose and 
Fermi parts
\bea
\int_0^1 ds \lrb{ x'_\mu(s) Q_{\mu\nu}\lrb{\pp{s}} x_\nu(s) +
\psi_\mu(s) Q_{\mu\nu}\lrb{\pp{s}} \psi_\nu(s)},\br 
Q_{\mu\nu}(\omega) =  \frac{\omega}{4T}  \delta_{\mu\nu} + \frac{i}{2} 
B_{\mu\nu},
\eea
with the only difference coming from the boundary
conditions.\footnote{Let us note in passing that the periodic boundary
conditions $\psi(1) = \psi(0)$ correspond to the chiral anomaly
instead of the vacuum energy. We discuss this issue later in some
detail.} 

Let us expand the fields in Fourier expansion on the circle
\bea
x_\mu(s) = \sum_{-\8}^\8 x_\mu^n \EXP{2 n \i \pi s},\br
\psi_\mu(s) = \sum_{-\8}^\8 \psi_\mu^n \EXP{(2 n + 1)\i \pi s},
\eea
then the functional integral yields the product of ratio of the 
determinants for each Fourier mode
\be
V \sqrt{\det Q(0)}\prod_{-\8}^\8 \frac{\sqrt{\det Q((2 n + 1)\i 
\pi)}}{\sqrt{\det Q(2 n \i \pi)}}.
\ee

The first factor corrects the contribution from the zero mode
$x_\mu^0$. There is, in fact, no Gaussian integration for this mode in
our action, so that this integration produces the volume $V$ of
space. The factor $\sqrt{\det Q(0)}$ removes the zero mode factor from
the denominator  at $n=0$ in the product. 

Now, the standard computation of the infinite product yields
\be
\label{EH}
V \int_0^\8 d T T^{\alpha-1}\lrb{1 + \alpha \ln T}e^{-T m_0^2}\sqrt{\det 
\lrb{B \cot 2 B T}}.
\ee
which is our representation of the Euler-Heisenberg Lagrangian. The
determinant of the matrix function is implied here. One could expand
it in power series in $B$(we denote the Bernoulli numbers as 
${\cal B}_{2n}$)
\bea
B \cot 2 B T = \frac{1}{2T}\lrb{1-\sum_{n=1}^\8 \frac{|{\cal 
B}_{2n}|}{(2n)!} (4BT)^{2n}},\br
\sqrt{\det \lrb{B \cot 2 B T}} = T^{-\oh d} \sum_{n=0}^\8 C_{2n}(B)  
T^{2n}, 
\eea
with some invariant polynomials
\be
C_{2n}(B) = \sum_{\{k\}} c_{\{k\}} \prod \lrb{\tr B^{2i}}^{k_i}.
\ee
The proper time integrals
\be
\int_0^\8 d T T^{\alpha-1}\lrb{1 + \alpha \ln T}e^{-T m_0^2}T^{2n -\oh
d} = \pp{\alpha}\lrb{\alpha m_0^{d - 4n - 2 \alpha}\Gamma\lrb{2 n  +
\alpha - \oh d}}. 
\ee
The pole at $d = 4$ in the $B^2$ term is the usual UV divergence,
removed by our $\alpha$ renormalization as follows 
\be
\pp{\alpha}\lrb{\alpha m_0^{-2\alpha}\Gamma\lrb{\alpha}} \ra -\gamma - 2 
\ln m_0.
\ee

\section{Supersymmetrization Theorem}

Let us consider the function $F\lrb{S_1,\dots S_n}$ of $n$ variables
$S_i = (s_i,\theta_i)$ which linearly transforms under infinitesimal
local SUSY transformations 
\bea
\delta s = \epsilon \theta \beta(s),\br
\delta \theta = \epsilon \beta(s).
\eea
The finite transformation have the form
\bea
s(\lambda) = s + \lambda \theta \beta(s),\br
\theta(\lambda) = \theta + \lambda \beta(s) + \oh \lambda^2 \theta 
\beta(s) \beta'(s).
\eea
One may verify that these polynomials satisfy differential equation
\bea
s'(\lambda) = \theta(\lambda) \beta(s(\lambda)),\br
\theta'(\lambda)= \beta(s(\lambda)).
\eea

The local SUSY implies that the equation $F\lrb{S_1(\lambda),\dots
S_n(\lambda)} = 0$ is, in fact, independent of $\lambda$ for arbitrary
Grassmann function $\beta(s)$.  Using this independence, we may choose
$\beta(s)$ so that it "passes through all the $\theta$ points"   
\be
\beta(s_i) =  \theta_i.
\ee
An explicit example of such function is a Lagrange interpolating polynomial
\be
\beta(s) = \sum_i \theta_i  \prod_{j\ne i} \frac{s-s_j}{s_i-s_j}.
\ee

Now, taking $\lambda = -1$ we observe that
\be
\theta_i(-1) = 0, s_i(-1) = s_i.
\ee
So, the function, which vanishes identically at the real axis, will
vanish identically in whole superspace. Therefore, one can extend in
superspace any invariant identity which holds on a real axis. 

Being valid at any finite $n$, in a limit it must be also valid for 1D 
functionals
$F\lbrack  \Gamma\rbrack$ which depend on some curve $\Gamma: S =
(s(\tau),\theta(\tau))$ in superspace. The same transformation with
$\lambda = -1 $ moves this curve to the real axis provided 
\be
\beta(s(\tau)) = \theta(\tau).
\ee
Any solution of for inverse function $ \tau(s) $ will give an example of 
such transformation 
\be
\beta(s) = \theta\lrb{\tau(s)}.
\ee

Such an obvious  theorem must, of course, be well-known to
mathematicians, but for a lazy physicist it was easier to "prove" it
than to dig into the mathematical literature. 

\section{Factorization of Fourier Integral}

The factorization of the Fourier exponentials
\be
\EXP{\i \int_A^B d P_\mu(U) X_\mu(U)} = \EXP{\i \int_A^S d P_\mu(V)
X_\mu(V)} \EXP{\i \int_S^B d P_\mu(W) X_\mu(W)}, 
\ee
follows from the  additivity of the $dX_\mu$ form. When we
compute the gradient precisely in a breaking point, in a complete
integral it is done by means of the integration by parts, with the
usual result 
\be
\EXP{-\i \int_A^B d P_\mu(U) X_\mu(U)}\pp{P_\mu(S)}
\EXP{\i \int_A^B d P_\mu(U) X_\mu(U)} = -\i d X_\mu(S).
\ee

In a split form one can obtain the same result in one plays by the
rules. This computation goes as follows 
\bea
-\i\EXP{-\i \int_A^S d P_\mu(V) X_\mu(V)}\ff{P_\mu(S)}
\EXP{\i \int_A^S d P_\mu(V) X_\mu(V)} =\br
\int_A^S d \delta(V-S) X_\mu(V)=
\delta(0) X_\mu(S) - \delta(A-S) X_\mu(A) - \int_A^S d 
X_\mu(V)\delta(V-S)=\br 
- \int_A^S d X_\mu(V)\delta(V-S) =
-\int d V D X_\mu(V) \delta(V-S)\Theta(A,V) \Theta(V,B) =\br
- D X_\mu(S)\Theta(A,S) \Theta(S,S) =
- \oh D X_\mu(S)\Theta(A,S).
\eea
The second exponential gives another half of the derivative $DX_\mu(S)$.

The same result could be obtained much simpler, using the Fourier $\oh$ rule
\be
D X_\mu(S) = \oh DX_\mu(S-) + \oh D X_\mu(S+).
\ee
This $\oh$ rule is a general rule to interpret the discontinuities. 

The terms with $S-$ ($S+$) is the boundary value of the ordinary
derivative, acting on the left (right) exponential, where they can be
replaced by functional derivatives $ \ff{P_\mu(S\pm)}$ for the inside
points.   
The left derivative $\ff{P_\mu(0-)}$ is equivalent to $
\ff{P_\mu(1)}$, so it acts on $\SC{P}_S^1$, whereas the right one acts
on $\SC{P}_0^S$. In the same way, the left derivative $
\ff{P_\mu(S-)}$ acts on $\SC{P}_0^S$, and the right one acts on
$\SC{P}_S^1$. As a result we get all four possible terms, each with
the weight $\oq$. 

\section{Generating Function for Super-Graphs}

There is a simple solvable model which serve as generating function
for these super-graphs. Consider the sum of two {\em Abelian} constant
gauge field strengths $G_{\mu\nu} + B_{\mu\nu}$ . The superpath
integral is computed in Appendix B, and it reduces to a Pfaffian 
\be
\int {\cal D} X \delta^d(X(0)) \EXP{\ih (G_{\mu\nu} + B_{\mu\nu}) \int_0^1 
d  X_\nu(S) X_\mu(S) }  \propto \sqrt{\det (B + G)}.
\ee

On the other hand, we could replace the $G$ simplectic form by an 
oscillator commutation relation
\be
\label{COM}
\left\lbrack  R_\mu,R_\nu\right\rbrack = \i G_{\mu\nu},
\ee
and compute the vacuum average instead
\bea
\int {\cal D} X \delta^d(X(0)) \vac{\TEXP{ \int_0^1 d X_\mu(U) R_\mu 
}}_R\br
\EXP{\ih B_{\mu\nu} 
\int_0^1 d X_\nu(S) X_\mu(S) } 
 \propto \sqrt{\det (B +  G)}.
\eea
This formula (see  \ct{Mig94}) follows from the definition of the the
ordered product and the Baker-Hausdorff formula. Another proof is to
switch to Schwinger gauge in the ordered exponential,
treating $R_\mu$ as a non-Abelian gauge field $R_\mu(S)$. This field
is constant, but it has a finite field strength \rf{COM}. This field
strength being Abelian, the Taylor expansion in a Schwinger gauge
terminates, so that one may replace 
\be
R_\nu(S) \Ra \ih G_{\mu\nu} X_\mu(S),
\ee
which recovers the initial Abelian formula.

Comparing the terms of expansion in powers of $G$ we get relations
\bea
\vac{R_{\mu_1}\dots R_{\mu_{2n}}}\int_0^1 \VEV{ d^{2n}
X_{\lbrace\mu\rbrace}(\lbrace U\rbrace)}_{Gauss} = \br
\lrb{\EXP{-\sum_{k=1}^{\8}\frac{(-1)^k}{2 k} \tr (B^{-1}G)^k }}_{n} ,
\eea
with the Gauss averages
\be
\label{GAUSS}
\VEV{DX_\mu(S) DX_\nu(S')} = \i B^{-1}_{\mu\nu} D\delta(S-S').
\ee
On the \RHS the term $\O{G^{n}}$ is left in the Taylor expansion.
The vacuum average can be computed from commutation relations, which
gives the set of sum rules for the locally SUSY integrals  we need.

The point is this set of relations is valid for {\em arbitrary $d$}, while
the SUSY integrals are universal numbers, {\em independent of $d$}.
So, these relations do not terminate at $ n = 4$ in four dimensions,
where we need them. 
 
Let us be more specific. Expanding the exponential on the right side, we get
\be
\sum_{l=0}^{n} \frac{(-1)^{n-l}}{l!} \sum_{k_1=1}^{n+1-l}\dots
\sum_{k_l=1}^{n+1-l} \delta_{n,\sum k_i} \prod_{i=1}^l \frac{\tr
\lrb{B^{-1}G}^{k_i}}{2 k_i}.
\ee
Each factor of $B^{-1}$ corresponds to the Gauss contraction
\rf{GAUSS}, and each factor of $G$ corresponds to the commutator
\rf{COM}. The indexes between the $B^{-1}$ and $G$ tensors are
contracted according to the specific Gauss diagram. The coefficient in
front of this product of tensors $B^{-1} \otimes\dots B^{-1} \otimes G \otimes\dots G$ on
the left is given by the SUSY integral
\be
\pm \int_0^1 d^{2n}U \prod_{\left\langle ij \right\rangle}
D\delta\lrb{U_i-U_j}.
\ee
The same coefficient on the right is given by above sum of $
\frac{(-1)^{n-l}}{l!}\prod_{i=1}^l\inv{2 k_i}$, independently of $d$.
In particular, the first relation reads
\bea
\oh \tr B^{-1} G  = -\i B^{-1}_{\mu\nu}\vac{R_\mu R_\nu} \int_0^1 d^2U
D \delta(U_1-U_2) =\br
-\oh \i B^{-1}_{\mu\nu}\vac{\COM{R_\mu}{R_\nu} }\int_0^1 d^2U
D \delta(U_1-U_2) = \br
\oh B^{-1}_{\mu\nu} G_{\mu\nu} \int_0^1 d^2U
D \delta(U_1-U_2) =  - \oh   \tr B^{-1} G \int_0^1 d^2U
D \delta(U_1-U_2),
\eea
from which we get
\be
\int_0^1 d^2U D \delta(U_1-U_2) = -1.
\ee

These relations in higher orders leave some undetermined SUSY integrals.

The typical integral involves the products of the $\Theta$ functions
and their derivatives, i.e. $\delta$ functions. The number of $\delta$
functions is twice less than number $2n$ of integration variables,
moreover, there are $\delta$ functions for every variable, otherwise
the local SUSY would be violated.  

The simplest graphs are those, when each $\Theta$ function is
differentiated only once. In this case all the points will collapse to
one of the endpoints of the interval $0,1$. Some $\Theta$ function
will collapse as well, leading to a bubble graph at one or another
endpoint,  

\input epsf \centerline{ \epsfbox{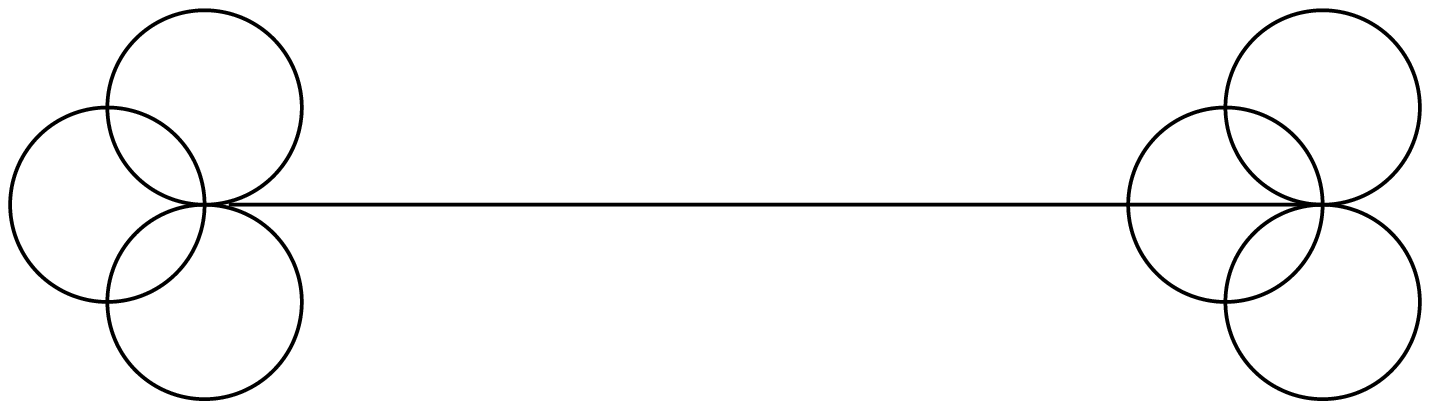}}
but some other will connect the endpoints, contributing just
$\Theta(0,1) = 1$. Each bubble will  give $\Theta(0,0) = \Theta(1,1)
=\oh$, so that the integral would be $\pm 2^{-k}$ where $k$ is the
number of bubbles. 

The generic term which arises in momentum expansion involves derivatives of the $\delta $ functions
\be
\pm \int_0^1 d^n S \int_0^1 d^m  K_{\lbrace \mu\rbrace}(\lbrace U\rbrace) \, \prod D \delta(U_b - S_c)
\ee
where $\lbrace 1\dots m\rbrace $ is partitioned into $\lbrace
a\rbrace,\lbrace b\rbrace$ and $\lbrace c\rbrace = \lbrace 1,\dots
n\rbrace$. 
There are precisely $n$ indexes in the $\lbrace b\rbrace$ list, and
$m-n$ in the $\lbrace a\rbrace$ list. 

Substituting here the above definition of $K_\mu(U)$ we get
superposition of terms with products of Cronecker deltas times 
\bea
\int_0^1 d^n S \int_0^1 d^m U \prod \delta(U_a - S_i) \, \prod D \delta(U_b - S_c)=\br
\int_{-\8}^\8 d S_1 \dots \int_{-\8}^\8 d S_n \int_{-\8}^\8 d U_1 \dots \int_{-\8}^\8 d U_m \br
\Theta(0,S_1)\dots \Theta(S_n,1) \Theta(0,U_1)\dots \Theta(U_m,1) 
\prod \delta(U_a - S_i) \, \prod D \delta(U_b - S_c)
\eea
These superspace integrals are universal numbers, which depend on the
partition of indexes. This follows from the local SUSY, which is
preserved here. 

The derivatives $D$ in the last line can be integrated by parts,
either over $S_c$ or over $U_b$ in such a way, that  they act only on
the $\Theta$ functions,  rather than $\delta$ functions. We get then 
\bea
\int_{-\8}^\8 d S_1 \dots \int_{-\8}^\8 d S_n \int_{-\8}^\8 d U_1 \dots \int_{-\8}^\8 d U_m 
\prod_l \Theta(S_l,S_{l+1}) \prod \Theta(U_k,U_{k+1})\br
\prod\delta(U_j -U_{j+1}) \prod\delta(S_r -S_{r+1})  
\prod \delta(U_a - S_i) \, \prod  \delta(U_b - S_c)
\eea

The number of $\delta$ functions equals $m + n$, therefore all the
integrals can be done. In case there are cycles,  there will be some
$\delta(0)$ factors. All such terms drop, as  
\bea
\delta(0)= 0,\br
\delta(S-S')\delta(S'-S) = 0,\br
\delta(S-S')\delta(S'-S'')\delta(S''-S) = 0,
\dots
\eea  
in superspace. So do the terms with the cycle of $\Theta$ functions,
as we already discussed above. These identities can be proven by
explicit calculation. 

So, only terms without cycles will be left. In those terms we shall
have the products like $\delta(U_i-U_{i+1})
\Theta(U_i,U_{i+1})$. These products can be replaced by $\oh
\delta(U_i-U_{i+1})$, since $\Theta(U,U) = \oh$.  


\begin{thebibliography}{99}

\bibitem{Mig94} A.A.Migdal, Second Quantization of the Wilson 
Loop,PUPT-1509, hep-th/9411100.
\bibitem{MM} Yu.M.Makeenko, A.A.Migdal, Exact equation for the loop
average in multicolour QCD, {\it Phys. Lett.}, 88B, 135-137, (1979) 
\bibitem{Mig83} Loop Equations and $\inv{N} $ Expansion, by A.A.Migdal, 
{\it Physics Reports}, 102, 199-290, 1983,
\bibitem{KazKos} V.A.Kazakov and I.K.Kostov, Nucl. Phys. {\bf B176}
(1980) 
199; Phys. Lett. {\bf B105} (1981) 453;
V.A.Kazakov, Nucl. Phys. {\bf 
B179} (1981) 283.
\bibitem{Cvitanovic}
CLASSICS ILLUSTRATED: GROUP THEORY. PART 1. 
By Predrag Cvitanovic (Nordita), Print-84-0261 (NORDITA), Jan 1984. 206pp.
Nordita Notes,
THE PLANAR SECTOR OF FIELD THEORIES. 
By Predrag Cvitanovic (Nordita), P.G. Lauwers (NIKHEF, Amsterdam), P.N.
Scharbach (Rutherford), NIKHEF-H/82-2, Jan 1982. 40pp. 
Published in Nucl.Phys.B203:385,1982.
\bibitem{Voiculesku} D. V. Voiculescu, K. J. Dykema and A. Nica, {\it Free 
Random Variables}, AMS 1992. 
\bibitem{Gross94} Rajesh Gopakumar, David Gross, {\it Mastering the Master 
Field} PUPT1520, 1994.
\bibitem{Mig86} Momentum Loop Dynamics and Random Surfaces in QCD, by 
A.A.Migdal, {\it Nucl. Phys.} B265 \lbrack  FS15\rbrack, 594-614, (1986),
\bibitem{BVM} Fourier Functional Transformations and Loop Equation. By 
M.A. Bershadski, I.D. Vaisburd, A.A. Migdal (Moscow, Cybernetics Council), 
1986. Yad. Fiz. 43 ( 1986) 503-513.
\bibitem{Pol87}  Gauge Fields and Strings, by A.M.Polyakov, {\it Harwood 
Academic Publishers} , 1987.
\bibitem{Tseytlin} O.D.Andreev, A.A.Tseytlin, Partition-Function
representation for the Open Superstring Effective action: Cancellation
of M\"{o}bius infinities and derivative corrections to Born-Infeld
lagrangian, Nucl.Phys. B311, 205-252, (1988). 
\bibitem{Luis} L.Alvarez-Gaum\`{e},  Commun. math. Physics 90(1983) 161; 
J.Phys A16 (1983) 1177;
\bibitem{Dan}D.Friedan and P.Windey, Nucl Phys. B235 (1984) 395.
\bibitem{Zuber} Claude Itsykson and Jean-Bernard Zuber, Quantum Field
Theory, Mc-Graw Hill, 1980, p. 195. 

\end{thebibliography}
\end{document}